\documentstyle[prl,aps,multicol]{revtex}
\begin{document}
\tightenlines
\newcommand{\be}{\begin{equation}}
\newcommand{\ee}{\end{equation}}
\newcommand{\bq}{\begin{eqnarray}}
\newcommand{\eq}{\end{eqnarray}}
\def\id{{\rm 1\kern-.21em 1}}
\title{Time-dependent mean field
theory of the superfluid--insulator phase transition}
\author{Luigi Amico$^{1,2}$ and Vittorio Penna$^{3}$}
\address{$^{1}$ Departamento de F\'{\i}sica Te\'orica
de la Materia Condensada $\&$ Instituto ``Nicol\'as Cabrera''
\\
Universidad Aut\'onoma de Madrid, E-28049 Madrid, Spain.}
\address{$^{2}$ Dipartimento di Metodologie Fisiche e Chimiche
per l'Ingegneria, Facolt\'a di Ingegneria,
Universit\'a di Catania \& INFM, viale A. Doria 6, I-95129 Catania,
Italy.}
\address{$^{3}$ Dipartimento di Fisica, Politecnico di
Torino \& INFM, Corso Duca degli Abruzzi 24, I-10129 Torino, Italy.}
\date{\today}
\maketitle
\begin{abstract}
We develop a time--dependent mean field approach, within the
time--dependent variational principle, to describe the
Superfluid--Insulator quantum phase transition.
We construct the zero temperature phase diagram  both of the
Bose--Hubbard model (BHM), and of a spin--$S$
Heisenberg model (SHM) with the XXZ anisotropy. The phase diagram
of the BHM indicates a phase transition from a Mott insulator to a
compressibile superfluid phase, and shows the expected lobe-like
structure. The SHM phase diagram displays a quantum phase transition
between a paramagnetic and a canted phases showing as well a lobe--like
structure.
We show how the BHM and Quantum Phase model (QPM) can be rigorously
derived from the SHM. Based on such results, the phase boundaries
of the SHM are mapped to the BHM ones, while the phase diagram of
the QPM is related to that of the SHM.
The QPM's phase diagram obtained through the application of our
approach to the SHM, describes the known onset of the
macroscopic phase coherence from the Coulomb blockade regime
for increasing Josephson coupling constant. The BHM and the QPM
phase diagrams are in good agreement with Quantum Monte Carlo
results, and with the third order strong
coupling perturbative expansion.
\end{abstract}
\begin{multicols}{2}

\section{Introduction}
Phase transitions induced by thermal or quantum fluctuations have
been studied in various mesoscopic systems. Examples are Josephson
Junction Arrays (JJA)~\cite{NATO}, granular~\cite{GRANULAR}, and
short--lenght superconductors~\cite{FILMS}. Such systems have 
two different critical temperatures $T_1$ and $T_0$ ($T_0 <T_1$).
Below $T_{1}$, they possess finite domains in which the electrons
form the Cooper pairs: In each domain the condensate is described
by the Cooper pairs' wave function $\psi_j = \Delta \, e^{i\Phi_j}$
($\Delta=|\psi_j|$ beeing related to the  pair density)
and the system is globally resistive
because of the absence of phase coherence between the Cooper
pairs~\cite{PAALANEN}. Below $T_0$, superconductivity may be
extended to a macroscopic scale (global superconductivity)
if the system reachs the macroscopic phase coherence.
\\
If the characteristic energy scale of the system is much smaller
than $|\Delta|$, one can regard the Cooper pairs as true
bosons~\cite{NOTE_G-S} and the global superconducting phase
transition can be studied by analyzing the critical behavior of
strongly correlated bosonic models on a lattice~\cite{BOSONIC,FISHER}.
These exhibit two characteristic energy scales: the hopping
amplitude $t$ which accounts for the boson kinetic energy,
and the Coulomb repulsion $U$ which is the electrostatic
energy expense to make bosons spatially close. The BHM can 
describe the energetic competition between $t$ and $U$.
The global superconducting phase transition is controlled by
the ratio $t/U$ which is a measure of {\it quantum effects}.
For $t \ll U$, the strong quantum fluctuations of $\Phi_j$
prevent the system from reaching the phase coherence for any
value of the temperature. The condition $t \gg U$ entails the 
classical regime: The system undergoes a phase transition at a
finite value of temperarture $T_0$ that belongs to the 
$D$-dimensional XY model's universality class.
Below $T_0$ the system is superfluid, while above $T_0$ it
becomes resistive.
If the Coulomb interaction is comparable with the kinetic energy
then quantum fluctuations make vanish $T_0$, and drive the ($T_0 =0$)
Superconductor--Insulator (SI) phase transition.
The latter has been studied in great detail both
experimentally~\cite{SI-EXP} and theoretically~\cite{BOSONIC}. The
superfluid phase is characterized by off-diagonal-long-range-order
signalled by a non vanishing order parameter $\Psi =\langle
e^{i\Phi_i} \rangle $ ({\it macroscopic quantum phase coherence}).
The insulating phase is incompressible and it is characterized by
$\Psi =0$.
In particular, due to the dimensional crossover, the SI phase
transition belongs to $(D+1)$--XY model's universality class
for commensurate bosons' densities, whereas it is mean
field--like away from such values~\cite{FISHER}.
\\
We recall that the number of bosons $n_j$ (at each site $j$)
is standardly considered to be canonically conjugated with $\Phi_j$.
This establishes a competition  between the quantum fluctuations of
$n_j$, and those of $\Phi_j$'s~\cite{NOTEFLUCT}

In a recent paper~\cite{AMPE} we formulated a Time--Dependent Mean
Field Theory (TDMFT) of the BHM in order to investigate some aspects
of the physical scenery just described. The TDMFT was based on
factorizing slow/fast dynamics described by an effective form
of the BHM Hamiltonian. The latter was derived within the
Time--Dependent Variational Principle (TDVP) procedure, and relied
on a picture of the system quantum state in terms of Glauber coherent
states.
In this approach the hamiltonian degrees of freedom
identified by construction with the parameters $z_j(\tau )$
--$\tau$ is the real time--of coherent states $|z_j\rangle$, that
is the expectation values ($\langle z_j |a_j |z_j \rangle$) of the
operators $a_j$ ($a^+_j$) describing at each site $j$ the destruction
(creation) of bosons~\cite{NOTEFT}. 
\\
We revealed that quantum effects
concerning the competition between the Coulomb term and the hopping
term are embodied in the time dependence of the coherent state
parameters $z_j(\tau)$.
The TDMFT involves a time--dependent, {\it local} order parameter
which is assumed to represent the slowly varying part of $z_j$, and
plays the same role of $\psi_j$.
In~\cite{AMPE}, we have shown that $\psi_j$ has a
time--independent amplitude which is the analogue of $\Delta$,
and a time--dependent phase which is the analogue of
$\Phi_j $. In particular, the phase's quantum
fluctuations were described in terms of phase's time--fluctuations.
\\
The phase transition is signalled by a qualitative variation of
the time behavior of the local superconducting order parameter.
In spite of the approximations involved by the TDMFT, indeed our
phase diagram shows a good agreement with Quantum Monte Carlo
(QMC) simulations~\cite{QMONTE} and Strong Coupling
Perturbative Expansion (SCPE)~\cite{FREERIKS}.

The purposes of this paper is to extend the TDMFT of the
BHM developed in Ref.~\onlinecite{AMPE} both to the SHM and to
the QPM for constructing their zero temperature phase 
diagrams\cite{UNIVCLASS}.
To this end we establish a rigorous mapping between the SHM, the
BHM and the QPM based on the Holstein-Primakoff Realization and
the Villain Realization of the spin algebra.
In particular, we shall see that the existence of such a mapping
is crucial to construct the phase diagram for the QPM
within the TDMFT of the SHM.
In this case in fact the explicit representation of the
Hamiltonian in terms of coherent states is
problematic~\cite{CARRUTHERS}
due to the euclidean algebraic structure of QPM's degrees of freedom.

In outline the paper is organized as follows. In Sec. II, we
introduce the three models we deal with, we illustrate some
basic aspects of their algebraic structure and describe their
qualitative phase diagrams.
%
%
%
Based on the TDMFT, we construct in Sec. III the SHM's phase diagram.
The latter will be shown to describe the SI transition only in some
interval of external magnetic fields. The discussion developed
here will concern the SHM's phase diagram only for magnetic field
in that range. The full description of the SHM's phase diagram will
be reported elsewhere~\cite{AMPESPIN}.
In Sec. IV we employ the program developed in Appendix A to recover
the BHM phase diagram from the SHM one, and to obtain the quantum
Josephson model's phase diagram from the SHM phase boundaries.
In Sec. V we give our conclusions and further remarks.
Appendix A is devoted to give a novel procedure following which both
the BHM and the QPM are obtained from the SHM.
In Appendix B we apply the TDVP method to work out the semiclassical
BHM Hamiltonian and its dynamical equations. Then we formulate the
TDMFT and employ it to construct the BHM's phase diagram.
After reviewing the basic properties of spin generalized coherent
states~\cite {MOPE}, the TDVP method is implemented in Appendix C
for the SHM.
In Appendix D we derive the phase dynamics of the QPM as
a perturbation of the SHM minimum energy configurations.

\section{The Models}

A convenient starting point for introducing models that exhibit
the SI phase transition is the BHM~\cite{NATO}. It represents a
boson gas of identical charges hopping through a $D$--dimensional
lattice whose Hamiltonian reads:
\be
H_{_{BH}}=\!
\sum_{j} \left [U ( n_{j} -1)-\mu \right ] n_{j}
- \! {{t}\over{2}} \sum_{\langle i,j\rangle}
\! \left (a^{\dagger}_{i} a_{j}+ a^{\dagger}_{j} a_{i} \! \right ) ,
\label{BHM}
\ee
\noindent
where the operators $n_{i} := a^{\dagger}_{i} a_{i}$ count the
number of bosons at the site $i$, and  the annihilation and creation
operators $a_{i}$, $a^{\dagger}_{i}$ obey the canonical commutation
relations $[a_{i},a^{\dagger}_{j}]= \delta_{ij}$. The set
$\{ {\bf \id}, a_{i}\, a_{i}^{\dagger}\, n_i \}$ is the basis 
generating at each site a Heisenberg-Weyl algebra $h_4$.
Also, the parameters $t$, $U$ of (\ref{BHM}) are the hopping
amplitude and the strength of the onsite Coulomb repulsion,
respectively, while the chemical potential $\mu$ 
fixes the average number of bosons in each site.
\\
The phase diagram of the BHM has been studied thoroughly by means
of mean field \cite{FISHER} and variational~\cite{VARIATIONAL}
approaches as well as perturbative~\cite{FREERIKS} and Quantum
Monte Carlo~\cite{QMONTE} techniques.
At $t/U=0$, the minimum energy configuration is characterized by
an integer number $n$ of bosons at each site, and a finite energy
gap $\mu=2 U$ for the creation of particle-hole excitations.
This reflects the Mott insulator (MI) behavior of such a phase
which entails a vanishing compressibility.
The MI regime survives (except for the degeneration points with
$\mu\! =\! 2nU$) when $t/U > 0$, inside extended {\it lobes}
attached to intervals $I(n)=(\, 2(n-1)U \, , \, 2nU \,)$ 
of the $\mu/U$ axis in the $t/U $--$\mu/U$ plane. 
Elsewhere, in the phase plane, the system exhibits a superfluid
character, both compressible and independent from the filling.
\\
At the lobe boundary the appearence of the
superfluidity is announced by the vanishing of the energy gap
between the states corresponding to $n$ (or $n-1$) and
$n+1$ (or $n$) particles (or holes). Also, at the critical points
the variance $\Sigma^2(\Phi)$ of the phase of the superconducting
order parameter is  reduced {\it so much} as the quantum coherence
can take place. Indeed, the $\Phi_i$'s quantum fluctuation survives
also in the superfluid phase and they are rigorously vanishing
($\Sigma_i^2(\Phi) =0$) only in the {\it classical limit}
$t/U \rightarrow \infty$. Such two features characterize the whole
MI--SF phase boundary as well as the onset of superfluid state.

The QPM is deeply related to the BHM. It is largely employed for
the description of quantum JJA's in which the phases $\Phi_i$ of
the superconducting order parameter are dynamically relevant,
the fluctuations of the modulus $\Delta$ being negligible
at low temperatures.
Since among the islands of nanofabricated samples no Ohmic
current flows and quasiparticle tunneling is actually negligible,
the QPM Hamiltonian defined as
\be
H_{_{QP}} =\! \sum_{ {i} , {j} }
Q_i \,V_{ij}\,Q_j  - \! {{E_{_J}}\over{2}} \sum_{<i,j>}
\cos \left (\Phi_{i}-\Phi_{j}  \right ) ,
\label{QPM}
\ee
with $Q_i := N_{i}-N_x$, can capture the physics of JJA.
In (\ref{QPM}), $N_{i}$ counts the number of Cooper pairs
which make the island deviate from the neutrality state with respect
to the background charge. The standard assumption that phase operators 
$\Phi_i$ are canonically conjugated to $N_i$'s, namely 
$[\Phi_{i},N_{j}]=2e i \; \delta_{ij}$, requires coherently 
that the $N_i$ eigenvalues must range from $-\infty$ to $+\infty$. 
The Coulomb interaction is described by the matrix
$V_{i,j}= 4e^2(C^{-1})_{i,j}$, where $C_{i,j}$ is the inverse of 
the capacitance matrix. In the sequel we will assume 
$V_{i,j}= V \delta_{i,j}$ with $V \equiv 4e^2 C_0^{-1}$,
where $C_0$ is the self--capacitance. The external 
voltage $V_x\doteq N_x/V$ enters via the induced charge $eN_x$, and
fixes the average charge on each island. The phase 
diagram of the QPM is similar to BHM's one (see \cite{OTTERLO,BALTIN}).
The MI lobes are attached to the interval
$I(N_0)= ((N_0-1/2)V ,\, (N_0+1/2)V)$, where the average
number of Cooper pairs is $N_0 \doteq int(N_x)$.
The degeneration points at $E_{_J}/V = 0$ are  $N_x = N_0/2 $.
Outside the lobes the system is globally coherent and exhibits a
superfluid character.

The argument usually employed to map the BHM on QPM is heuristic:
In the limit of large average number of boson per site the bosonic
field may be represented as $ a_j \simeq \sqrt{ n_j } e^{i \Phi_j }$,
$a^\dagger_j \simeq e^{-i\Phi_j}\sqrt{n_j}$ ($\Phi_j$ being hermitian),
where the operators $n_j\doteq N_j-N_x$ have negligible fluctuations,
and play the role of Cooper pairs' density. The QPM is recovered
through the identifications $N_x \equiv \mu/\sum_{j} U_{i,j}$ and
$E_{_J} \equiv n t$ ($n$ is the average boson density).
Since the fluctuations of $n_j$ are underestimated in
the QPM the superfluid region is smaller than the BHM one.
\\
It is worthwhile noting how a rigorous reading of this mapping leads
to severe inconsistencies~\cite{CARRUTHERS}.  These are mainly due to
the boundness from below of bosonic number operators $n_j$ that makes
extremely complex
the problem of defining the hermitian phases
canonically conjugated to $n_j$~\cite{NOTECAN}.
Nevertheless, the QPM can be defined in a consistent way
since the operators $N_i := -i \partial_{\Phi_i}$, and
$\exp{(\pm \Phi_i)}$ with $\Phi_i$ an {\it hermitian phase}
are employed in formula (\ref{QPM}).

The difficulties involved in mapping the BHM into the QPM in a direct
way can be circumvented by using a third (spin) model as a bridge
connecting the first ones. To this end we first describe the standard
way to relate anisotropic spin--$1/2$ Heisenberg model with the 
physics of the SI transition, then we introduce the spin--$S$ model 
that will be employed in the sequel.
\\
At very low temperature, few charge states are important if on-site
Coulomb repulsions are very large. If the gate voltage is tuned close
to a degeneracy, two charge states per island actually suffice to
represent the relevant physics.
Therefore the {\it hard core} BHM is equivalent to a spin-$1/2$
Heisenberg model~(see Ref.~\onlinecite{LIU}) the model Hamiltonian 
of which is represented in terms of $su(2)$ operators
$ S^{z}_{i}, \; S^{+}_{i},\;S^{-}_{i}$ in the
fundamental representation.
Substitutions $S^{z}_{i} \to n_{i}- 1/2$, $S^{+}_{i}
\to a^{+}_i$, and $S_i\to a_i$ allow one to recover the
hardcore BHM from the spin--$1/2$ Heisenberg model.
\\
The zero temperature phase diagram of the spin--$1/2$ Heisenberg
model has been investigated in Ref. \cite{LIU,MATSUDA} within
the mean field approximation. This shows the presence of a
phase transition between paramagnetic/Neel phases in which
$\langle S^z_i \rangle \! \neq \! 0 $ and
$\langle S^\pm_i \rangle \! = \! 0$, and
a {\it canted} state in which
$\langle S^z_i \rangle = \langle S^\pm_i \rangle \! = \! 0$.
\\
Such magnetic orderings have a counterpart in the BHM's phases:
the canted state (long range order in $\langle S_{i}^x \rangle $
and $ \langle S_{i}^y \rangle $)
indicates superfluidity in the BHM; the paramagnetic/Neel phases
(long range order in $\langle S_{i}^z\rangle$:
$\mid \! \langle S_{i}^z \rangle \! \mid \! \neq \! 0 $) correspond
to the MI's (the role of $\mu$ is played by the Heisenberg model
external magnetic field $h$).
The representation in terms of the (anisotropic) spin-$1/2$ Heisenberg
model cannot work for the soft core BHM, this involving,
in general, more than two charge states .

We propose the spin--S XXZ Heisenberg model (denoted so far by SHM)
as a model capable of describing the physics of the SI
phase transition when more than two charge states are involved.
In this case the Hamiltonian has the form of the anisotropic spin
$1/2$ Heisenberg model, but (the representation index) $S>1/2$.
It reads
\begin{eqnarray}
H_{{_S}} = \sum_{i} \, (S^{z}_{i} +S) [&U&(S^{z}_{i}+S-1) - h]
\nonumber \\
&-& {{E_{_S}}\over{2}} \sum_{\langle {i} , {j} \rangle}
\left (  S^{+}_{i} \; S^{-}_{j} +
 S^{+}_{j} \; S^{-}_{i} \right ) \, ,
\label{SHAM1}
\end{eqnarray}
where we have taken into account only the on-site
interaction along $z$. In Appendix A we describe a procedure 
which maps the SHM onto the BHM and the QPM.
%
%
\section{TDMFT of the SHM}
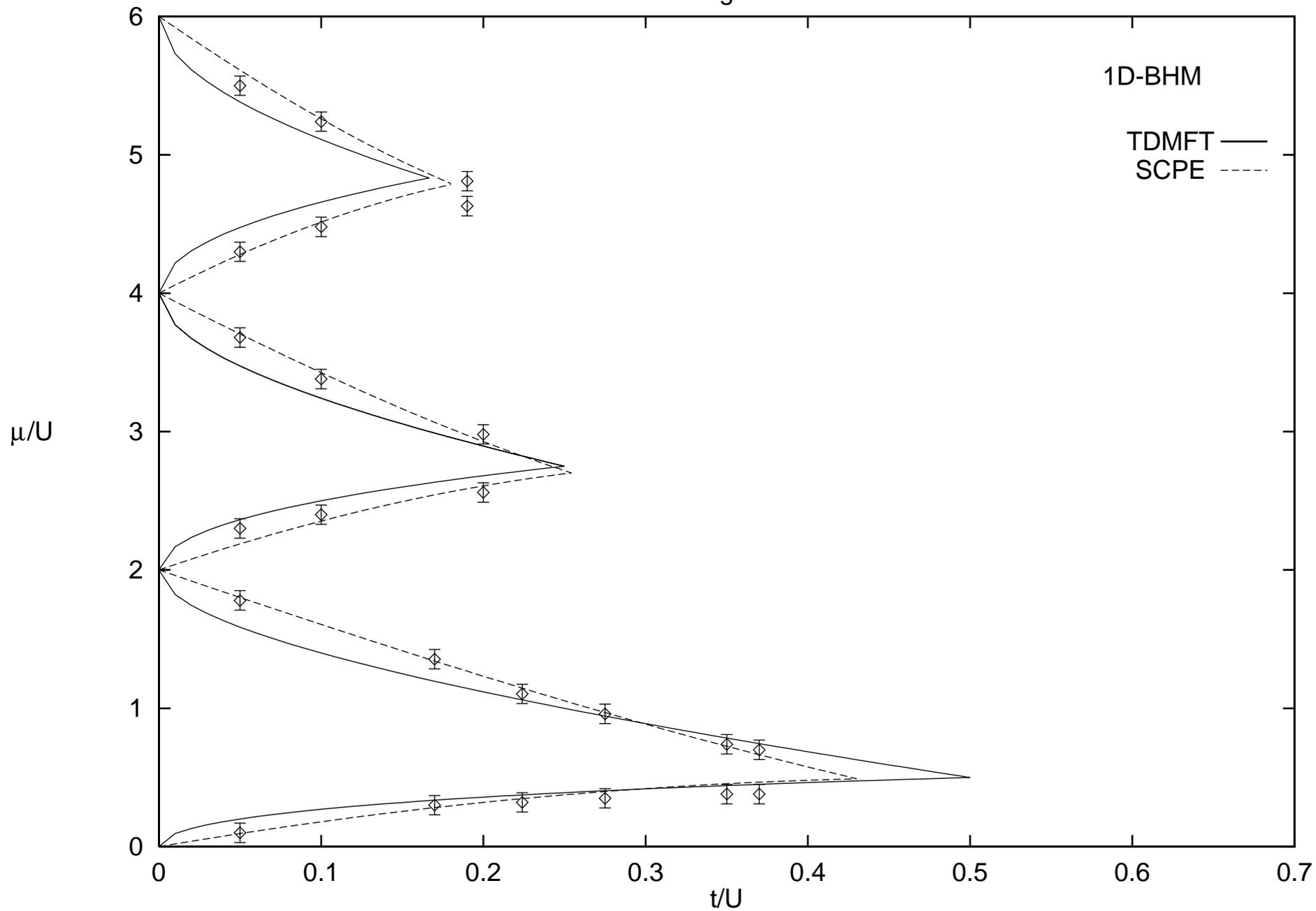
\begin{figure}
\caption{The phase diagram of the  $BHM$ for $D=1$. The
error boxes are the QMC results of {\it Batrouni et al.} in Ref.[9].
The dashed lines are the result of the SCPE.
Relatively to the first lobe ($n_{i}=1$), Eq.~(\ref{TIP}) gives 
$(t_{c}/U)=0.5$.
QMC gives $(t_{c}/U)=0.43 \pm 0.002$ and the SCPE
$(t_{c}/U)=0.43$.
For ($n_{i}=3$), QMC and SCPE give  $(t_{c}/U)=0.2$ and
$(t_{c}/U)=0.18$ respectively. Our theory gives $(t_{c}/U)=0.16$.}
\label{BHMD1}
\end{figure}

In this section we construct the phase diagram representing the $SI$
phase transition for the SHM and compare it with that of the BHM
represented in Figs.~\ref{BHMD1},\ref{BHMD2}.
We develop a construction the main
steps of which are very similar to the TDMFT of the BHM
depicted in Appendix B; thus we will sketch them without
comment when the analogy with the TDMFT of the BHM is evident.

The semiclassical SHM model is achieved by projecting Hamiltonian
(\ref{XXZM}) on the $su(2)$ coherent states (see Appendix B),
thereby obtaining
\be
{\cal H}_{{_S}} = \! \sum_i ( U\,L^{z}_{i} - h_* ) L^{z}_{i}
- \frac{E_{_S}}{2} \sum_{\langle {i}, {j} \rangle} \!
\left (L^{*}_{i} L_{j} + L^{*}_{j} L_{i} \right ) +C,
\label{XXZM}
\ee
where $h_* :=h-U(2S-1)$, and $C := \! \sum_i S[h_* +U(S-1/2)]$.
The semiclassical equations of motion obtained within the TDVP 
method~\cite{ZFG} read
\begin{equation}
i \hbar {\dot L}_j= \left [U (2 L^z_j +2 S-1) -h \right ] L_j
+ E_{_S} L^z_j \sum_{i\in (j)} \! L_i \, .
\label{SEESPIN}
\end{equation}
Eqs. (\ref{SEESPIN}) are not integrable, since the constant of 
motion available are just two, namely the $z$--component of the 
total angular momentum $L_z =\sum_j L_j^z$ and the energy itself.
The TDMFT decoupling simplifies Eq.~(\ref{SEESPIN}) through the
approximation
\begin{equation}
L_{i} \; L^{*}_{j} \approx
{\cal M}_i L^{*}_{j}+ L_i {\cal M}^{*}_j -{\cal M}_i {\cal M}^*_j
\label{LINEARSPIN}
\end{equation}
which ensues from $(L_i-{\cal M}_i)(L^*_j -{\cal M}_j)\approx 0$ and
entails $\langle L_j \rangle_{\tau}={\cal M}_j$ on long time scales
(compare with~(\ref{MEAN}) and~(\ref{ORD})). The order parameter is
defined as
\begin{equation}
M \equiv {\frac {1}{N_s}} \sum_j \langle L_j \rangle_{\tau} \, .
\end{equation}
When (\ref{LINEARSPIN}) is inserted in~(\ref{XXZM}) together with
the uniformity condition ${\cal M}_j\equiv {\cal M} \; \forall j$,
Hamiltonian~(\ref{XXZM}) reduces to a sum of on-site terms:
${\cal H}_{_S} \rightarrow \sum_j {\cal H}_j$, where
\begin{eqnarray}
\label{XXZMF}
{\cal H}_j =  const &+& (U\,L^{z}_{i} -h_* )L^{z}_{i} \\
&-&{{q E_{_S}}\over{2}}
\left (  {\cal M}_j L^{*}_{j}+ {\cal M}_j^{*} L_{j}-|{\cal M}_j|^2
 \right ) \;. \nonumber
\end{eqnarray}
The equations of motion derived from~(\ref{XXZMF}) simplify to
\begin{equation}
i \hbar {\dot L}_j= \left ( 2U L^z_j - h_* \right ) L_j
-q E_{_S} L^z_j {\cal M}_j
\label{EQXXZ}
\end{equation}
and imply that $L_z$ is no longer a constant of motion.
In analogy to the BHM case of Appendix B, implementing the
phase--locking condition $\phi_j -\alpha_j = \{ 0, \pi \}$
on $L_j=|L_j|e^{i \phi_j}$ and
${\cal M}_j = |{\cal M}_j| e^{i \alpha_j}$,
($\phi_j$, $\alpha_j \in [ 0, 2\pi ]$) successfully
restores the basic feature $dL^z/dt=0$.
Due to Eqs. (\ref{EQXXZ}) the phases $\phi_j$
obey the equation
\begin{equation}
-\hbar L_j {\dot \phi}_j =
\left (2U L^z_j - h_* \right ) L_j- q E_{_S} L^z_j {\cal M}_j \, .
\label{XXZPHI}
\end{equation}
We examine first the spin dynamics related to the paramagnetic--like
phase. Such a phase is identified by $\langle S^z_j \rangle$ sharply
picked at one of its spectral values $m$: $\Sigma^2 (S^z_i) \ll 1$;
this condition (due to the uncertainty principle) induces strong
quantum fluctuations of $S^x_j$ and $S^y_j$ which suppress the
ferromagnetic order in the $x-y$ plane.
Semiclassically, $\Sigma^2(S^z_i) \ll 1$ translates in $L^z_j = m$.
\\
Within our scheme, the condition $L^z_j = m$, $\forall j$ follows
from the requantization procedure~\cite{GUZ,IKG} of the actionlike
variables $L^z_j$ (notice that $\{ L^z_j ,\phi_j \}=
\delta_{ij}/ \hbar$), and entails $|L_j| = \sqrt{S^2-m^2}$.
Since Eqs.~(\ref{XXZPHI}) is solved by
$\phi_j = \tau \rho_{\pm}/\hbar + \phi_j(0)$, then
\begin{equation}
{\cal M}_j= \pm {{U \Delta+ \rho_{\pm}}\over{q E_{_S} L^z_j}} L_j \, .
\label{SELFCON}
\end{equation}
We have parametrized the external magnetic field as
$\Delta :=  h_* /U-2m $.
The parameter $\rho_{+}$ ($\rho_{-}$) is related
to the choice $(\phi_j-\alpha_j)=0$ ($(\phi_j-\alpha_j)=\pi$).
The order parameter
$$
M = \frac{1}{N_s} 
e^{i\tau \rho_{\pm}/\hbar}\sum_j {\cal M}_j e^{i \phi_j(0)}
$$
has a vanishing (long) time average $\langle M \rangle=0$, because
of the time-dependent phase factor $e^{i \tau \rho_{\pm}/\hbar}$.
Such a phase factor forbids the breaking of the original $so(2)$
symmetry of the SHM in the $xy$ plane. Thus, the oscillating behavior of $M_j$
identifies the paramagnetic phase. Notice that the condition
$\langle  M \rangle=0$ can be realized also for ${\cal M}_j\neq 0$:
As in the case of the TDMFT of the BHM, the TDMFT of the SHM
can describe the paramagnetic phase for $E_{_S} >0$.

The frequencies $\rho_{\pm}$'s play the role of
time correlation lenght governing the phase transition.
The observations of the previous section
as to the criticality of BHM can be extended to
model~(\ref{SHAM1}): As expected, the critical exponents $z$
and $\nu$ fulfill
the same Eq.~(\ref{EXP}). The latter is left unchanged by the 
procedure mapping the SHM onto the BHM and QPM. Hence, 
this suggests that the same Eq.~(\ref{EXP}) governs the 
criticality of the QPM.

Eqs~(\ref{SELFCON}) play the role of the self--consistent
equations of the TDMFT: They serve to eliminate
the order parameter from the energy~(\ref{XXZMF}).
The energy of the paramagnetic phase reads:
\begin{eqnarray}
\label{ENPARA}
{\cal E}_m (h,E_{_S} ;\rho)&& = -U (\Delta + m+S) (m+S)  \\
&-&{{2 q E_{_S} m-\rho_{\pm} -U \Delta }\over{2 q E_{_S} m^2}}
\left [U \Delta + \rho_{\pm} \right] (S^2\! -m^2) \, . \nonumber
\end{eqnarray}
The condition $h_*/U -2 m =0$ ($\Delta=0$) identifies the
degeneration points at which the canted phase reachs the axis
$E_{_S}/U=0$. The points on such axis correspond to static solutions
of the equation of motions~(\ref{EQXXZ}) which are obtained through
$\rho_{\pm}=0$. Such a condition is the simplified form of the low
frequency dynamics in the canted state.

Now, we consider the fixed points of the dynamical equations
(\ref{EQXXZ}) that identify the canted phase. In such a case
the self-consistent equations read
\begin{equation}
{\cal M}_j={{ h_* -2 U L^z_j}\over{q E_{_S} L^z_j}} L_j
\label{SELFCONCAN}
\end{equation}
The calculation of the energy minimimum of the canted phase
is considerably  simplified by applying the
change of variable
\begin{equation}
p_j={{ h_*-2 U L^z_j}\over{q E_{_S} L^z_j}} \, ,
\end{equation}
which implies that the self-consistency equation can be written as
\begin{equation}
{\cal M}_j = p_j
\sqrt{S^2-\left ( {{ h_*}\over{q E_{_S} p_j+2 U}}\right )^2} \, .
\end{equation}
After noting that the order parameter ${\cal M}_j$ is a monotonic
(increasing) function of $p_j$ we are allowed to eliminate the order 
parameter from the energy~(\ref{XXZMF}) then minimizing it with
respect to $p_j$. The energy reachs its minimum value at $p_j=1$
which corresponds to
\begin{equation}
(L^z)_{min}={{h_*}\over{q E_{_S} +2 U}} \equiv {\cal L}_0 \;,
\label{JMINCANT}
\end{equation}
where the index $j$ has been dropped since the uniformity of the
solution.
We point out that inserting $p_j=1$ in~(\ref{SELFCONCAN}) implies
${\cal M}_j= L_j$ on which the TDMF decoupling is based.
The minimum energy is
\begin{equation}
{\cal{E}}_{min}= U\frac{S}{2} -{{(h+qS E_{_S} +U)^2}\over{2 (q E_{_S} +2 U)}}
\label{ENMINCANT}
\end{equation}
Energy (\ref{ENMINCANT}) gives the known (see \cite{MATSUDA}) 
classical value of the energy minimum (up to the constant $US/2$, 
see Appendix B). 
Thus we conclude that the static solutions of
Eqs.~(\ref{EQXXZ}) correspond to the classical canted phase.
\\
Now we employ the energy~(\ref{ENPARA}) to obtain the phase
boundaries between the paramagnetic and the canted phases.
The curves representing the $m$--phase boundary are
identified  by implementing the $so(2)$ symmetry breaking
through the limits $\rho_\pm\rightarrow 0$ and
the vanishing of the energy gaps ${\cal E}_m -{\cal E}_{m \pm 1}$
(compare with Eqs.(\ref{REQ1}), (\ref{REQ2})). In particular,
${\cal E}_m ={\cal E}_{m +1}$ provides the equation
\begin{eqnarray}
{{U}\over{2 q E_{_S}}} \Delta_{-}^2 -
r_m \,\Delta_{-} + s_m =0
\label{DELTA}
\end{eqnarray}
where $\Delta_{-}\doteq h/U-2 (m+S) \le 0$, and
$r_m := m(m+1)/(1+2 m)$,
$s_m :=  \left [2(m+S) +1\right ] (1+2m) r_1^2/S^2$.
A second equation is derived from ${\cal E}_m ={\cal E}_{m -1}$
where $\Delta_-$ must be replaced by 
$\Delta_{+}\doteq h/U-2 (m+S) +2 \ge 0$.
As in the bosonic case, the upper (lower) root issued from
${\cal E}_m ={\cal E}_{m -1}$ (${\cal E}_{m+1} ={\cal E}_{m}$)
give the phase boundaries
\end{multicols}
\begin{equation}
{{h_+}\over{U}}=2 (m+S-1) - {{q E_{_S} m (m+1)}\over{U (2 m+1)}}
\left[-1+
\sqrt{1-{2U\over{q E_{_S} S^2}}
(2 m+1)(2 (m+S)+1)}\right ]
\label{XXZPDP}
\end{equation}
\begin{equation}
{{h_-}\over{U}}
=2 (m+S)- {{q E_{_S} (m+1) (m+2)}\over{U (2 m +3)}}\left[-1-
\sqrt{1-{2U\over{q E_{_S} S^2}}
(2 m+3)(2 (m+S)+3)}\right ]
\label{XXZPDM}
\end{equation}
\begin{multicols}{2}
We study the phase diagram~(\ref{XXZPDP}), (\ref{XXZPDM}) for
$m<-1/2$. Such a restriction
guarantees that the canted phase propagates
up to $E_{_S}/U=0$ at the degeneration points $\Delta=0$ for any
value of $m<-1/2$ (otherwise the paramagnetic lobes meet one each
other at $\Delta = 0$, but at $E_{_S}/U \ne 0$).
This feature is a common property of the phase diagrams which
describes the SI transition. The above condition selects values
of the external magnetic field $h$ in the range $-2 U<h<(S-3/2) U$.
Such a limitation will be relaxed in a forthcoming
paper~\cite{AMPESPIN}.
\\
Fig.~\ref{S-XXZ} shows the phase diagram of the SHM for $S=10$.
We note that the coordinates $(E_{_S} /U)_{tr}$ of the tricritical
points (the cusps of the lobes) are not monotonic as function of $m$
(see also Fig.~\ref{TIPS}).
In the lower part of the phase diagram they {\it decrease}
monotonically (increasing $m$ starting from its minimum value
$-S$, $h = -2$ in Fig.~\ref{S-XXZ}) up to a critical value of $m$ at which
they are almost independent on $m$ (we shall call ``inversion region''
such a portion of the SHM phase diagram).
In the lower part of the phase diagram, the shape of lobes is
asymmetric. Instead, within the inversion region, the lobes are
almost symmetric around odd--integer values of $m+S$.
Increasing $S$ makes expecially wide  the inversion regions.
In such a region we can assume the effects of the $L^z$'s dynamics 
independent on $m$.  
For larger values of $m$, $(E_{_S} /U)_{tr}$ increases
monotonically with $m$.
Fig.~\ref{TIPS} shows the non-monotonic behaviour 
of the tricritical points for various values of $S$.  
In line with the mapping developed in Sec. III, in the next
section we show that the SHM phase diagram contains the BHM
(QPM) phase diagram in the lower (inversion) region. 

\section{From the SHM to the BHM and QPM phase diagram}
In this section we apply the results of section III to construct
the phase boundaries of both the BHM and the QPM
from the phase diagram of the SHM.

We first recover the BHM phase diagram as a limiting case of the
phase boundaries~(\ref{XXZPDP}), (\ref{XXZPDM}) for $S\gg1$,
$\alpha \rightarrow 0$. The latter allows one to expand the energy
of the paramagnetic phase~(\ref{ENPARA}) in power of $\alpha$;
upon relating the parameters $E_{_S}$, $h$ with $t$, $\mu$ as in
section III A, we find that the energy of the paramagnetic phase
(\ref{ENPARA}) reduces to the form
\end{multicols}
\begin{equation} \label{ENEREXPA}
{\cal E}_m (h,E_{_S};\rho_\pm) = E_n (\mu , t;\rho_\pm)
- n \left [1+ 2 \; {{\rho_{\pm}+U(\delta +1)}\over{q t}}\right ]
\left [\rho_{\pm}+U(\delta+1)\right ] \alpha + {\cal O}[\alpha^2]
\; .
\end{equation}
\begin{multicols}{2}
Such a formula shows that, at the zero-$th$ order in $\alpha$,
${\cal E}_m $ matches the on-site energy in the Mott phase~(\ref{X1}).
In Fig.~\ref{XXZvsBHM} we compare the phase diagram of the BHM
with the SHM phase diagram worked out using the energy~(\ref{ENPARA})
for $S=55$.

Similarly, we can obtain the phase diagram of the QPM by taking
the limit $S, m \gg 1$, $\beta^2 \ll 1$ of the phase boundaries
(\ref{XXZPDP}), (\ref{XXZPDM}). As pointed out previously, we recall
that a coherent states picture of QPM is not yet available in terms
of coherent states of $e(2)$  (which is the algebra
of the QPM microscopic degrees of freedom, see (\ref{E2}))~\cite{CARRUTHERS}.
Hence, the direct application of the TDMFT, which relies on the
coherent states description of the hamiltonian operators,
cannot be implemented in a direct way.
The mapping outlined in section III B is crucial
to bypass such difficulties. It allows, in fact,
to construct the QPM phase diagram within the SHM lobe-like
structure provided the values of $m/S$ remain inside a 
suitable range (see below).
As stated in Sec. III B, the condition $S \gg |m|$ together with the
identifications $U \equiv V$, $N_x \equiv h_* /2U$, and
$E_{_J} \equiv S^2 E_{_S} (1-\beta^2)$ recast the SHM Hamiltonian in the 
QPM form (see appendix C). Correspondingly, the phase boundaries 
take the form
\end{multicols}
\begin{equation}
N_x^\pm={{2 m \mp 1}\over{2}}+
{{q \beta}\over{U S}}
\left [ E_{_J} \mp \sqrt {E_{_J}^2+8 {{UE_{_J}}\over{q}} S^2
\beta (1-\beta) }\right ] \, ,
\label{QPMPB}
\end{equation}
\begin{multicols}{2}
We note that for sufficiently large $S$ small changes of $m$
leave the parameter $\beta$ almost unaltered. This is sufficient
to make the SHM phase diagram periodic in $N_x$.
The curves~(\ref{QPMPB}) give a qualitatively correct QPM phase
diagram for any $\beta^2\ll 1$.
We fix $\beta$ in~(\ref{QPMPB}) to reproduce quantitatively the
QMC phase diagram. The Fig~\ref{QPMPD} shows the phase diagram
of the SHM model for $S=55$ and $\beta^2=0.028$. 
The paramagnetic phases of the SHM turn in to the Coulomb blockade
of the Cooper pairs. The canted state reveal the macroscopic phase
coherence of the QPM. 

\section{Conclusions}

In this paper we have been concerned with two different aspects
of the SI transition. The former is the algebraic structure that
characterizes the main models exhibiting the SI phase transition.
The latter is the development of the Time-dependent mean field theory
(TDMFT) for the spin $S$ Heisenberg model with XXZ-anisotropy (SHM)
that generalizes the approach previously elaborated for the Bose
Hubbard Model (BHM). Remarkably, such a theory appears to be
applicable to a large class of spin models.

The analysis of appendix A concerning the algebraic framework in
which the SHM is constructed, indicates that the correct way to map
the SHM on the BHM is given by the Holstain-Primakoff realization of
spin operators. Instead, the Quantum Phase Model (QPM) is related
to SHM through Villain's realization of the spin operators. 
Thanks to the transparent geometric meaning of such realizations
\cite{MAT,IKG} both the BHM and the QPM can be issued from the
SHM by considering the appropriate sectors of the spin spectrum. 
To summarize: the BHM Hamiltonian emerges in the limiting case
of spin vectors close to the south pole of the spin sphere,
while the QPM behavior is found for spin vectors around the
sphere equatorial plane.

The TDMFT have been presented in an extensive way in appendix B,
where it is used to investigate the BHM and its SI
phase transition. Such a theory is based on combining the
Time-dependent variational principle (TDVP), the coherent state
picture of the model quantum dynamics, the Einstein requantization
procedure, and a time-dependent generalization of the mean field
decoupling. 
\\
The central assumption in our theory is that at $T=0$, time
dependence of the semiclassical variables $z_i(\tau)$ represents
the analogue of the quantum fluctuations of operators $a_i$.
\\
Although Eqs. (\ref{SEE}) (the semiclassical counterpart of
Heisenberg's equations for $a_j$'s) have been simplified to describe
an on--site dynamics concerning $z_j$'s, they provide a consistent
description of the system's quantum phase transitions. This is due
to the semiclassical requantization procedure (see~(\ref{REQUANT})),
on the one hand, and on the time dependence characterizing the phase
of the superconducting order parameter, on the other.
The latters make the order parameter phase the main
responsible in driving the SI transition. 
Within the TDMFT, we have established a relation between the
dynamical behavior of the (local) superconducting order parameter
and the macroscopic phases exhibited by the BHM. The Mott phase 
has been shown to be characterized by time fluctuations of the
phase of the superconducting order parameter, whereas the
superfluid phase is related to the static solutions
of the (mean-field) equations of TDVP.
The energy minimum coincides with the classical $t/U \gg 1$
superfluid case. 
\\
Below, we compare the ordinary MFA (Mean-Field Approximation) with
the TDMFT. The first comment in order is that the time-independent
MFA of the BHM~\cite{FISHER,SHE} describes the SI phase transition
by means of the suppression of the {\it amplitude} of the order
parameter. 
The quantum fluctuations of the phase of the superconducting order
parameter play no role in the standard MFA. Moreover, the MI phase
is characterized by an on-site energy where the hopping term
does not contribute.
\\
The main difference between TDMFT and the standard MFA stands
in the dynamical content of the definition of the superconducting
order parameter: Within the TDMFT, the SF phase is suppressed by 
the (order parameter's) {\it phase}'s time fluctuations just as
phase quantum fluctuations destroy macroscopic quantum coherence.
Also, in the MI phase only the time average of the local parameters
$\psi_j$ is vanishing (along large time scale). This does not imply
that $\psi_j $ is strictly zero inside the insulator as it happens
in the standard MFA.
\\
The good agreement with Quantum Monte Carlo  simulations and
Strong Coupling Perturbative Expansion is
{\it a posteriori} confirmation that the superfluid phase is
almost classical: Quantum fluctuations are strong in the insulating
phase; they drive the SI phase transition, and are negligible in
the superfluid phase, except in the regions very near to the phase
boundaries. 

Based on the introduction of the spin coherent states performed
in Appendix C, the TDVP effective picture of quantum dynamics and
the TDMFT of Appendix B have been extended to the SHM in Sec. III.
Our analysis has revealed a quantum phase transition between a
paramagnetic and a canted phase.
The phase diagram exhibits a lobe-like structure. Inside the lobes 
the phase is paramagnetic; elsewhere the system is in a canted state. 
The energy in the canted phase
(represented by the stationary solutions of the equations of motions)
coincides with the known classical energy.
\\
We point out how the ordinary MFA (linearization of the $xy$ exchange 
term in (\ref{SHAM1})) leads to study the reduced (on--site)
Hamiltonian in the $su(2)$ enveloping algebra (due to the $({S^z})^2$
term) which prevents one from diagonalizing the Hamiltonian for generic
$S$~\cite{RASETTI}. On the other hand, fixing the representation index
leads to study the matrix--Hamiltonian whose rank increases with $S$. 
Thus, studying the spectrum of~(\ref{SHAM1}) for sufficently high $S$
(which is done in the present paper) would be very problematic.
Otherwise the classical analysis (see f.i.~\cite{MATSUDA}) cannot 
capture the lobe--like structure of the SHM phase diagram.

The SHM's phase diagram contains the BHM's and QPM's ones. Using the
strategy developed in Sec. III we have recovered the BHM energy as well
as its phase diagram. As we pointed out previously, implementing the
TDMFT to the QPM is problematic since the semiclassical description
of the model in terms of the $e(2)$ coherent states is not available.
Nevertheless, the phase diagram of the QPM has been obtained as an
appropriate limit (see appendix A and IV) of the SHM phase diagram
(the lobe structure is (locally) periodic around $m$ for changes
$m +\Delta m$ such that $\Delta \! m /S \ll m/S$). This picture
appears to be consistent with the QPM form assumed by the SHM
(effective) Hamiltonian close to its ground-state configuration
(see Appendix D), and suggests that possibly the (purely
quantum) SHM can be reduced to the QPM form as well.

Other perspectives are opened by the present study.
\\
First, the tools we used to map the SHM on the BHM and the
QPM show how the enveloping algebra characterizing the BHM
and QPM have common root in the enveloping algebra of the SHM. 
The mapping between these models can be seen as {\it contraction}
procedure~\cite{GILMORE} of the enveloping algebras underlying
the BHM, SHM and QPM~\cite{AMICO}. This suggests the fact that the
universality class might be preserved by contraction.
\\
Second,
since the equations of motion (of the BHM and the SHM ) obtained
by the TDVP have been considered within the simplified form
entailed by TDMFT, the dynamical approach
refined out the TDMFT should imply a more accurate description
of the superconducting order parameter dynamics as well as a
better understanding of the low temperature regime.
As to this point, other improvements can be achieved by
constructing trial wave functions able to account
more accurately the microscopic physical processes
(see~\cite{VARIATIONAL}).
Finally, the TDMFT succeeds in capturing the main features of the
quantum phase diagram of a rather large class of models.
Promising applications of the TDMFT to other systems
are expected due to its feasibile character.
%
\acknowledgments
The authors would like to thank L. Casetti, U. Eckern, G. Falci,
R. Fazio, R. Franzosi, G. Giaquinta, R. Maciocco, A. Osterloh, 
M. Rasetti and S. Sharov for valuable discussions. 
L.A. acknowledges financial suppport from EU TMR
Programme (ERB 4061 PL 95--0670), and the warm hospitality
of the Theoretical Physics II in Augsburg. V. P. expresses his
gratitude to the Schr\"odinger Institute in Wien, where part of
this work was done, for supporting his visit as well as to the
MURST for financial support within SINTESI Project.
%
%
%
%
\begin{appendix}
%
\section{Model Mapping}
In this appendix we give a new proof of the equivalence between the
SHM, the BHM, and the QPM. The key operation of such a mapping 
consists in studying the model~(\ref{SHAM1}) for 
sufficiently high values of representation index $S$,
and in using the Holstein-Primakoff
Realization (HPR) and the Villain Realization (VR)
of the spin algebra.
\subsection{From the SHM to the BHM}
We rewrite spin operators in Hamiltonian (\ref{SHAM1})
by means of the HPR of the spin algebra $su(2)$
\begin{eqnarray}
S^+_i&=&\sqrt{2S} a^{\dagger}_i\sqrt{1-n_i/(2S)} \nonumber \\
S^-_i&=& \left (S^+_i \right )^\dagger\nonumber \\
S^z_i&=& n_i-S
\end{eqnarray}
and note how, in view of the formulas of Appendix B,
the eigenvalues of the secular equation for the $i$-th spin
$(S^z_i + S)|m_i, S \rangle =(m_i +S) |m_i,S\rangle $ ranging
in $\{ 0,\dots , 2S \}$ identifies with the eigenvalues of $n_i$.
Consistently, the spin states $|m_i, S \rangle$ coincide with
the number operator eigenstates $|n_i \rangle$ up to the
reparametrization $n_i = m_i +S$.

The condition $n_i/S\equiv (m_i+S)/S=\alpha \ll 1$ allows
one to obtain Hamiltonian~(\ref{BHM}) from (\ref{SHAM1}).
In particular, the $xy$--exchange term of (\ref{SHAM1}) gets
in the hopping term of~(\ref{BHM}) with the hopping amplitude
$t= 2 (1-\alpha ) S E_{_S} \to \, 2S E_{_S}$ for
$S \rightarrow \infty$, and $n_i$ finite.
In the same limit, the spectrum of $S^z_i+S \equiv n_i$ ranges in
$\{ 0,\dots ,\infty \} $ thus reproducing the spectrum of bosonic 
operators $n_i$. The  $z$--antiferromagnetic term and the
coupling with the external magnetic field term becomes the Coulomb
interaction and chemical potential terms repectively ($h=\mu$).
\\
Also, since $\alpha \equiv (m_i +S)/S$ ($\alpha$ has to be viewed
as an order of magnitude independent on the site label $i$),
the above limit corresponds quantum mechanically to select spin
states close to lowest weight vector $|S,-S \rangle$ of the
algebra $su_i (2)$~\cite{NOTELIM1} at the $i$-th site.
In this respect, we recall that, within the
HPR, $|S, -S \rangle =|0 \rangle$, where $a_i|0 \rangle =0$.
\\
The effect of taking $\alpha \ll 1$ is illustrated
by means of the semiclassical spin vector $\vec L$ 
(defined in Appendix B) lying on a sphere of radius $S$.
Such a limit leads to select those vectors neighbouring
to the south pole of the sphere.
Consistently, spin coherent states $|\xi_i \rangle$ (see
definition (\ref{SCS})), that pertain to the Hilbert space
of bosons within the HPR, can be shown to tend to the bosonic
ones $|z_i \rangle$ for $\alpha\rightarrow 0$ (see (\ref{trial})),
while the same limit makes the spin vectors on the sphere coincide,
via stereographic projection, with the points of the south pole
tangent plane \cite{BPS}.
These are in bijective correspondence with the complex numbers
$z_j$ labeling the bosonic coherent states.
In fact, formulas (\ref{LZ}), (\ref{LR}) clearly show that
$L^z_j \to -S$ is achieved for negligible $|\xi_j |^2$ which
also entails $L_j^* \to 2S \xi_j$, $L_j \to 2S \xi_j^*$.
This in turn leads to the identifications
$z_j^* \equiv L_j^* /\sqrt{2S} = \sqrt{2S} \xi_j$,
$z_j \equiv L_j /\sqrt{2S} = \sqrt{2S} \xi_j^*$,
and makes coincide brackets (\ref{ALG}) with brackets (\ref{BOS}).
Hence the limit described above for the hamiltonian operators is
consistently reproduced at the classical level. We shall see in
Sec. VI that the SHM phase diagram matches the BHM phase 
diagram for $\alpha\rightarrow 0$.
%
%
\subsection{From the SHM to the QPM}
We write the spin operators in Hamiltonian~(\ref{SHAM1}) by means
of the Villain realization of $su(2)$ \cite{MAT}. This is based on 
the formulas
\begin{eqnarray}
S^+_j &=& e^{i\phi_j}\sqrt{(S+1/2)^2 -{{(S^z_{j}+1/2)^2}}} \, ,
\nonumber \\
S^-_j &=& \left (S^+_j \right )^\dagger \, .
\nonumber \\
\label{VIL}
\end{eqnarray}
Such operators fulfil the $su(2)$ commutation rules provided
the action-angle operators $S_j^z$, $\phi_j$, satisfy the $e(2)$
commutators
\begin{equation}
\left [S_{\ell}^z , e^{\pm i \phi_j} \right ]= \pm \, \delta_{j,\ell}
\, e^{\pm i \phi_j}\, , \, 
\left [ e^{i \phi_j} , e^{-i \phi_{\ell}}  \right ]= 0 \, .
\label{E2}
\end{equation}
The QPM~(\ref{QPM}) is obtained as a limiting model from
Hamiltonian~(\ref{SHAM1}) when considering the first order 
in $\beta = |m_i|/S \ll 1$ for $S \gg 1$.
The ferromagnetic part of ~(\ref{SHAM1}) reduces to the Josephson term
with coupling $E_{_J} \equiv (S+1/2)^2 E_{_S} (1-\beta^2)$ which
in the limit $S\rightarrow \infty$ becomes $E_{_J} \equiv S^2 E_{_S} $.
As in the case of $\alpha$, the parameter $\beta=|m_i|/S$ must be
regarded as a site-independent order of magnitude. Consistently,
the spectrum of $S^z_i$ will range from $-\infty$ to $+\infty$ thus
reproducing the spectrum of the unbounded operator $N_i$ in
(\ref{QPM}).
The rest of Hamiltonian~(\ref{SHAM1}) maps to the charging
term of~(\ref{QPM}) provided $V\equiv U$, and
$N_x=[h+U(1-2S)]/2U$. It results $H_{_S} \to H_{_{QP}} + C_0$,
where $C_0 = -(U+h)^2/4U$. 
\\
Geometrically, $\beta \ll 1$ amounts to select vectors neighboring
the equatorial $xy$--plane of the semiclassical sphere of radius $S$.
This is well illustrated via formulas (\ref{LZ}), (\ref{LR}) that
provide $|\xi_i|^2-1 \ll 1$ ($|L_i^z| \ll S$) as counterpart of the
above condition $\beta \ll 1$. Consistently, the sphere equation
$(L_i^z)^2 +|L_i^*|^2 =S^2$ shows that $|L_i^*|^2 \simeq S^2$.
Since both the Josephson coupling
and the hopping amplitude contain the factor $E_{_S}$,
the formula $E_{_J} = t\, S/2$ holds for $1 \ll S$, $\alpha \ll 1$,
$\beta \ll 1$. Thus $S$ plays the role of the boson density $n$
(see Sec. II). 
We point out that the effective hopping coupling in the BHM is
reduced by a factor $S$ when compared with the Josephson coupling
in QPM; consistently, the SF region of the BHM's phase diagram is
smaller than the SF region of the QPM's one.

We note how, when considering the perturbative
dynamics around the ground-state configuration of the SHM (see
Appendix C), one obtains a QPM-like behavior
(that is, having a pure phase's dynamics) without performing
the limit $S \to \infty$. We shall see that the phase diagram
of the QPM is obtained for $\beta^2\ll 1$~\cite{NOTELIM2}.

\section{The TDMFT of the BHM}

In this appendix we apply the TDVP (see Refs.\cite{ZFG} and
\cite{MOPE} for a general review) to the quantum dynamics of
Hamiltonian (\ref{BHM}), and implement the TDMFT for the BHM. 

\subsection{Time-dependent Mean field theory}

The initial step of the TDVP method amounts to finding
a solution of the Schr\"odinger problem
$(i\hbar \partial_{\tau} - H)|\Psi \rangle =0$ by approximating
the exact (unknown) solution $|\Psi \rangle $ through a macroscopic
state $|\Phi \rangle$ whose time evolution is imposed to obey the
weaker form of Schr\"odinger's equation
$\langle \Phi| (i\hbar \partial_{\tau}-H) |\Phi \rangle =0$.
Upon setting $|\Phi \rangle =exp({i{\cal S}/ \hbar}) |Z \rangle$
one obtains
\begin{equation}
{\dot {\cal S}}
= i\hbar  \langle Z| \partial_{\tau} |Z \rangle -
{\cal H} (Z)
\label{azione}
\end{equation}
\noindent
(${\cal H} (Z) \equiv \langle Z| H |Z \rangle$) 
which represents the key equation of the approach. 

The building blocks of Hamiltonian (\ref{BHM})  are operators of the
$N_s$-site Heisenberg--Weyl algebra $h_4(N_s)= \{ {\bf I}, a_{i},
a_{i}^{\dagger}:\,i\in \Lambda \}$, $N_s$ is the number of sites
of the lattice $\Lambda$, but actually belongs to the enveloping
algebra $\cal A$ of $W(N_s)$ because of the quadratic terms $n^2_j$.
This motivates the choice $|Z \rangle := \otimes_j |z_j \rangle$
as the trial macroscopic state, entailing
\be
|\Phi \rangle \equiv e^{i{\cal S(\tau)} / \hbar} \otimes_{i}
|z_{i}\rangle \,,
\label{trial}
\ee
where the states $|z_{i}\rangle$ are the Glauber coherent states
fulfilling the secular equation $a_{i}|z_{i}\rangle =
z_{i} |z_{i}\rangle$ for the boson lowering operator $a_{i}$, at
each site $i$. In this case the effective Lagrangian (\ref{azione})
becomes
\be
{\dot {\cal S}}[Z] =
i\hbar \sum_i {\frac {1}{2}}
({\bar z}_i {\dot z}_i - {\dot {\bar z}}_i z_i )
- {\cal H} (Z) \, ,
\label{action}
\ee
where ${\cal H} (Z) = \langle Z| H |Z \rangle$ -the semiclassical
model Hamiltonian- is given by
\begin{equation}
{\cal H} = \sum_{i} (U |z_i|^2 -\mu ) |z_i|^2 -
{{t}\over{2}} \sum_{\langle i,j\rangle}
\left ( {\bar z}_{i} z_{j} + {\bar z}_{j} z_{i} \right )\, .
\label{SBHM}
\end{equation}
\noindent
The equations obtained variationally from (\ref{action})
\begin{equation}
i \hbar \dot {z_i} =-\mu z_i +2 U z_i |z_i|^2 -
{{t}\over{2}}\, \sum_{j \in (i)} z_j \, .
\label{SEE}
\end{equation}
account for the dynamics of variables (expectation values)
$z_{i}=\langle z_i|a_{i}|z_{i}\rangle$.
Eqs. (\ref{SEE}) describe an hamiltonian flow in that they can be
equivalently obtained through the standard formulas
$i \hbar {\dot z_j } = \{ z_j , {\cal H} \}$,
where the Poisson Brackets
\be
\{ f(Z,{\bar Z}) , g(Z,{\bar Z}) \}
= {\frac{1}{i \hbar}} \!
\sum_j \! \left(
{\frac{\partial f}{\partial z_j}}
{\frac{\partial g}{\partial {\bar z_j}}} -
{\frac{\partial g}{\partial z_j}}
{\frac{\partial f}{\partial {\bar z_j}}} \right)\, ,
\label{BOS}
\ee
specifically, $\{z_k ,{\bar z}_j \} = \delta_{kj} /i\hbar$ have
replaced the basic commutators $[a_{i},a^{\dagger}_{j}] =\delta_{ij}$
within the TDVP semiclassical framework.
Eqs. (\ref{SEE}) are not integrable, since the only known constant of
motion, apart from the energy, is ${\cal N}=\sum_i |z_i|^2$, i.e. the
semiclassical version of the total particle number $N =\sum_i n_i$.
The presence of the nonlinear $U$--dependent term prevents one from
decoupling them in the dual lattice space.
\\
The TDMFT procedure is, in a sense, the analogue in a dynamical contest
of the Mean Field Approximation (MFA) usually employed in Statistical
Mechanics and is based on a well known microscopic picture of
superfluids illustrated, e.g., in Ref. \cite{AHNS}. It leads to
simplify the structure of Eqs. (\ref{SEE}).
We set at each site $z_i =\psi_i+\eta_i$, where $\psi_i$ is a slow
variable, whereas $\eta_i$ is a fast oscillating term which describes
the high-frequency part of the dynamics taking place on the hopping
interaction time scale. Also, we assume
that $(z_i - \psi_i)({\bar z}_j - {\bar \psi}_j) =
\eta_i {\bar \eta}_j \approx 0$.
Thus $\psi_j \equiv \langle z_j \rangle_{\tau}$ when the time scale
$\tau$ is larger than that of the $b_j$'s
($\langle \bullet \rangle_{\tau }$ denotes the time average). 
Such time averages coincide with statistical averages (in the
Gibbs ensemble) under the ergodic assumption.
The onset to the (macroscopically)  ordered  phase reflects the
presence in the system of stable, slowly varying components of the
lattice dynamics corresponding to the $\psi_j$'s. This means that
any $z_j$ is strongly attracted to its average value $\psi_j$
(namely that the collection of $\psi_j$'s defines the dynamical
system's attractor). Dynamical regimes where the long scale-time
behavior of $z_j$ is not described by an asymphtotic slowly
varying function $\psi_j$ is related with the disordered phases
of the system.
The above considerations imply the TDMF  decoupling
\begin{eqnarray}
z_i {\bar z}_j &&\equiv (z_i - \psi_i)({\bar z}_j - {\bar \psi}_j)
+\psi_i {\bar z}_j +
{\bar \psi}_j z_i - \psi_i {\bar \psi}_j\nonumber \\ &&
\approx \psi_i {\bar z}_j +
{\bar \psi}_j z_i - \psi_i {\bar \psi}_j \; .
\label{MEAN}
\end{eqnarray}
The dynamical scenery just depicted leads thus naturally to defining
\be
\Psi \equiv {\frac {1}{N_s}} \sum_j \langle z_j \rangle_{\tau}
\label{ORD}
\ee
as the macroscopic order paramenter revealing when order issues from
the the lattice dynamics. Using formula (\ref{MEAN}) in $\cal H$
modifies the kinetic term as follows
\be
{{t}\over {2}} \sum_{\langle i,j\rangle} \left ( {\bar z}_{i} z_{j}+
{\bar z}_{j} z_{i} \right ) \rightarrow
{{q t}\over{2}} \sum_{i}\left ({\bar z}_{i} \psi_{i}
+ {\bar \psi}_{i} z_{i}
-|\psi_{i}|^2 \right ) \, ,
\label{SUB}
\ee
where $\psi_{j} \equiv \psi_{i}$ for $j \in (i)$ (smoothing
condition). The resulting Hamiltonian reduces to the decoupled
form ${\cal H}_{mf} = \sum_j {\cal H}_j$, where
\be
{\cal H}_j = U|z_{j}|^4 - \mu |z_{j}|^{2} - {{q t}\over {2}}
\left ( {\bar z}_{j} \psi_{j} +{\bar \psi}_{j}  z_{j} -
|\psi_{j}|^2 \right ) \, ,
\label{SH}
\ee
and exhibits a dimensionality dependence entering via the numbers
of nearest neighbours $q$. The (decoupled) equations of motion
ensuing from ${\cal H}_{mf}$ read
\be
i \hbar \dot {z_{i}}=-\mu z_{i} +2 U z_{i} |z_{i}|^2 -
{\frac {qt}{2}} \psi_{i} \,.
\label{MFE}
\ee
When compared with the exact ones, (\ref{SEE}), they
imply the relation $q \psi_{i} \approx \sum_{j \in (i)} z_j$
consistently leading to an identity once the time average is
carried on and the smoothing conditions is used. A further effect
coming from the linearization (\ref{SUB}) consists in the fact that
the total particle number ${\cal N}=\sum_{i} |z_{i}|^2$ does not
have any longer vanishing Poisson brackets with ${\cal H}_{mf}$.
Restoring such a basic feature is performed by considering $z_j$
with an appropriate time dependence. To this purpose we look for
solutions of Eqs.~(\ref{MFE}) where $\theta_j$, $\chi_j$, the
phases of
\be
z_j = |z_j| \, e^{i\theta_j} \, \, , \, \, \,
\psi_j =|\psi_j| \, e^{i\chi_j} \, ,
\label{FUNC}
\ee
respectively, obey the phase locking condition
$(\theta_j -\chi_j)=\{0, \pi\} $. Then  ${\cal N}$
is constant due to the fact that
$d|z_j|^2/dt =iqt \, (z_j {\bar \psi}_j -{\bar z}_j \psi_j )/2 =0$.
Moreover, the further condition $d|\psi_j|/dt=0$ consistently makes 
${\cal H}_{mf}$ constant as expected for the total energy.
\\
Due to Eq. (\ref{MFE}), the phase $\theta$ obeys the equation
\be
- \hbar |z_j| {\dot \theta}_j = (2U|z_j|^2 -\mu) |z_j| -
s {{qt}\over{2}} |\psi_j| \,\,,
\label{TETA}
\ee
where $s= \pm$, depending on how the phase locking constraint is
implemented. In spite of its simplicity, such an equation is able
to characterize both the MI phase and the superfluid phase in terms
of phase dynamics.

We examine first the dynamics related to the MI. In this case,
$\psi_j$ must have a zero time average along macroscopic time scales.
Such a behavior occurs when the uniform filling conditions $n_j = n$,
for all $i$ (we identify here number operators $n_j$'s with their
integer spectral values) is inserted in (\ref{TETA}) by setting
\be
|z_j|^2 = n \in {\cal N} \, .
\label{REQUANT}
\ee
Such a substitution is the natural consequence of the requantization
process~\cite{GUZ,IKG}
of the action-like variables $|z_j|^2$ (notice that
$\{ |z_i|^2 ,\theta_j \}= \delta_{ij}/ \hbar $) strongly requested
from the pure quantum character of the MI.
At $t/U =0$, where the system is integrable 
(since it reduces to a set
of uncoupled, nonlinear oscillators) indeed $\theta_j$ and $|z_i|^2$
represent the pairs of action-angle variables of the system. For small
values of $t/U$ such a feature is still true as a consequence of the
fact that the nonlinear oscillators are weakly interacting in the MI
regime. Hence, in the spirit of Einstein's requantization procedure
(see \cite{GUZ}), their orbits are still homotopic to those of the
integrable case which entails again $|z_j|^2 = n$. 
\\
Eq. (\ref{TETA}) is easily showed to be solved by
$\theta_j(\tau) = \lambda_\pm \tau / \hbar + \theta_j(0)$,
with $\lambda_{\pm}$
defined through
\be
|\psi_{j}|\doteq \pm   {{2 \sqrt{ n }}\over{q t  }}
\left ( \lambda_{\pm } - U \delta     \right ) \; .
\label{LAMBDA}
\ee
Here we have parametrized the chemical potential as
$\delta = \mu/U -2 n $, and $\lambda_{-}$ ($\lambda_{+}$)
is related to the choice  $\theta_j -\chi_j =\pi$ ($\theta_j
-\chi_j =0$) (notice that the index $j$ does not label
$\lambda_\pm$ since the request $\langle z_j \rangle_{\tau} \equiv
\psi_j(\tau)$ leads to $|\psi_j| = \sqrt n$ at each site).
In the present theory, the frequencies $\lambda_{\pm}$
play the role of time correlation lenght
governing the phase transition.
Our theory gives
$\lambda_{\pm} = U \sqrt{n} (\mu-\mu_c )$ for fixed  $t$ and
$\lambda_{\pm} =q |\psi_{i}| /2 (t-t_c)$ for  fixed $\mu$
($\mu_c$ and $t_c$ are the critical values of $\mu$ and $t$).
Upon defining the critical exponents $z$ and $\nu$ as in
Ref.~\onlinecite{FISHER}, we argue that~\cite{COMMENT}
\be
z \nu=1 \; .
\label{EXP}
\ee
\\
By replacing  in the reduced Hamiltonian (\ref{SH})
the value of $|\psi_{i}|$ provided by
Eq. (\ref{LAMBDA}), the energy of the MI reads
\be
\! E_n \! (\mu ,t; \lambda_\pm)\! = \! n \left[ {\frac{2}{qt}}
(\lambda_\pm + U \delta )^2 \! +\!
U(\delta -\! n) -\! 2\lambda_{\pm} \! \right] ,
\label{X1}
\ee
where the subscript $n$
reminds us that the filling $n$ is accounted for.
The oscillating behavior of
$\Psi =(e^{i \tau \lambda_\pm } /N_s )
\sum_j \psi_j e^{i \theta_j(0)}$,
having a vanishing long time average, identifies the MI.
This, in fact, implies that the gauge symmetry breaking
expected in the SF phase cannot take place.
Notice that the ordinary (time independent) MFA cannot describe the
MI for $t>0$, since the hopping term of the reduced Hamiltonian
is canceled  by the vanishing of the order parameter, $\psi=0$.
Within our scheme, instead, the condition
$\langle \Psi \rangle_\tau =0$ can be realized also
for $\psi \neq 0$.
The degeneration points selected by $\mu/U=2n$ (i.e. $\delta =0$ in
$I(n)$) are extreme limiting points for which the superfluid phase is
extended  up to $t/U=0$. They will be identified with the meeting
points of the lobe boundaries. Such points characterize a static
phase due to  $\lambda_{\pm}=0$ (see Eq.~(\ref{TETA}).
We interpret the stationarity which distinguishes the solutions of
the semiclassical equation of motion as the trait characterizing the
SF phase in which $\Sigma^2(\theta)=0$ (classical SF). This
is but the oversimplified version of the low frequency dynamics
expected in the SF phase that should correspond
to the condition $\Sigma^2(\theta)\ll 1$.
\\
Let us consider the fixed points of dynamical equations~(\ref{MFE})
that, as we concluded above, identify the classical SF phase.
Such solutions (the trivial case $\dot z_j =0$ due to
$z_j = \psi_j =0$ is excluded) allow us to recast
Eq.(\ref{MFE}) in the form
$ \psi_j =\! 2[(2U|z_j|^2 -\! \mu)/qt] z_j$
making $\psi_j$ a function of $z_j$. Inserting $\psi_j (z_j)$
in~(\ref{SH}) reduces the energy associated
to the Hamiltonian ${\cal H}_j$ to
\be
\epsilon (\mu , t, z_j) \! = \! |z_j|^2 \!
\left[ {\frac{2}{qt}}  (\mu -\! 2U|z_j|^2 )^2
+\! \mu - \! 3U|z_j|^2  \right] .
\label{ENS}
\ee
The quantity $\epsilon (\mu , t, z_j)$ is the on-site energy
accounting for the absence of dynamics.
The limit $\lambda_{\pm} \rightarrow 0$, in fact, shows that
$ E_n (\mu , t;\lambda_\pm)  \rightarrow  \epsilon (\mu , t, z_j)$
provided $ n=|z_j|^2 $.
The lowest value of energy~(\ref{ENS}) and the value
$z_*$ involved for $z_j$ are obtained by minimization. They are
given by
\be
\epsilon_{*} =-U |z_{*}|^4 = - {{(\mu + qt/2)^2 }\over{4U}}\, ,
\label{ESC}
\ee
\be
|z_{*}|^2 = {{(\mu + qt/2)}\over{2U}}\, ,
\label{ZSC}
\ee
respectively.  The phase of $z_{*}$ can be  set to zero  since  the
gauge symmetry breaking characterizing the ground-state configuration.
It is worth noting that inserting $|z_{*}|$ in the expression
$\psi_j (z_j)$ implies that $\psi_j \equiv z_j$ so that the minimum
energy configuration naturally fulfils the consistency condition on
which our TDMFT is based.

Now, we employ the expression (\ref{X1}) for the on-site energy to
construct the BHM's phase diagram. In the SF phase, the states with
$n$ and $n+1$ (adding a particle), as well as the states with $n-1$
and $n$ (adding a hole) must be degenerate.
The curves representing the $n$-lobe boundary are
identified  by implementing both gauge symmetry breaking
through the limits $\lambda_\pm\rightarrow 0$ and  the
vanishing of the energy gaps $E_n - E_{n \pm 1}$. In other
words we require
\begin{equation}
\lim_{\lambda_{+} \rightarrow 0}\left (E_n - E_{n+1} \right)
=0 \; \makebox{\hspace{0.5cm}} \left (\delta <0 \right ) \; ,
\label{REQ1}
\end{equation}
\begin{equation}
\lim_{\lambda_{-} \rightarrow 0}\left ( E_{n-1}-E_n \right )=0 \;
\makebox{\hspace{0.5cm}} \left (\delta >0 \right ) \;.
\label{REQ2}
\end{equation}
For solving Eqs. (\ref{REQ1})-(\ref{REQ2})
we introduce the variables $\delta_{\pm} =\mu /U -2n+(1\pm 1)$.
By inserting $\delta_+ \ge 0$ ($\delta_-\le 0$) in eq. (\ref{REQ1})
((\ref{REQ2})), and  defining $r = qt/4U$, one gets the quadratic
equations $\delta_\pm^2 + 2 r \delta_\pm -(2n\mp1)=0$,
that furnish the pair of two-branched curves
\be
{\frac {\mu_\pm} {U}}
= 2n\! - \!1 \mp 1- \! {\frac{q t}{4 U}} \left [1 \mp
\sqrt{ 1 +  {\frac{8 U}{q t}} (2n\mp1)}\,\right ] \, .
\label{BB1}
\ee
The lower branch  $\mu_+ (t)$ and the upper one $\mu_- (t)$
constitute the boundary encircling the $n$-th lobe.
The substitution of $\mu_-(t)$ and $\mu_+(t)$ in (\ref{REQ1}) and
(\ref{REQ2}) respectively, provides the on-site energy values involved
in the two cases, namely $E_{n-1}(\mu_1,t) = E_{n}(\mu_1,t) = Un(n-1)$
and $E_{n}(\mu_2 ,t) = E_{n+1}(\mu_2,t) = Un(n+1)$. The two branches
are therefore separated by an energy gap. Thus,  the lobe tips
are singular points of the  energy. Their coordinates, obtained
imposing $\mu_-(t) = \mu_+(t)$, reads
\be
t_c = U/ qn
\label{TIP}
\ee
and $\mu(t_c)/U = 2n-1 -(1/2n)$. The values of $t_{c}$
obtained within the present theory, can be compared with
QMC and SCPE~\cite{FREERIKS} (Figs.~\ref{BHMD1},\ref{BHMD2}).
Contrary to the result obtained in Ref.~\cite{FREERIKS},
our phase diagram has a concavity independent of the dimensionality.
In 1D we find a good agreement with QMC and SCPE.
Upon recalling that our construction relied on Eq. (\ref{MFE}) --this 
incorporates the time-dependent mean field approximation-- it
is important to note that the concavity of lobes in $D \ge 2$
might be improved by implementing the requantization process
directly on the Eqs. (\ref{SEE}).
%
\subsection{Remarks on the SF ground-state}

As to the effectiveness of the approach just illustrated,
two important observations are in order. The first is
that our finding (\ref{ESC}), (\ref{ZSC}) concerning the ground-state
configuration is remarkably confirmed by two other procedures.
The other concerns the macroscopic phase at the phase transition (see 
Eq.(\ref{ANGSF}),(\ref{ANG})).
 
The minimum energy of the SF phase can be
calculated from (\ref{SBHM}). It represents the exact value of
the BHM ground-state energy in the classical
limit $t/U \gg 1$. 
The minimum energy is readily obtained by rewriting first the
hopping terms as $ ({\bar z}_{i} z_{j} +\! {\bar z}_{j} z_{i}) =\!
|z_{i}|^2 + \! |z_{j}|^2-|z_{j} -\! z_{i}|^2 $,
and by noticing then that the choice $z_{\ell} = \xi$ for each site
entails the lowest value of the hopping term since the only positive
contribution $|z_{j} - z_{i}|^2 $ vanishes on each bond. Upon
minimizing the resulting expression of the energy
\begin{equation}
{\cal H_*} = N_s \left [ (U |\xi|^2 - \mu) |\xi|^2
-{{tq}\over{2}} |\xi|^2 \right ] \, ,
\end{equation}
by setting $d{\cal H_*}/d|\xi| =0$, one obtains
$|\xi|^2 \equiv (\mu + qt/2)/2U $ that matches exactly (\ref{ESC}):
The ground-state energy (\ref{ENS}) 
coincides with the minimum of the exact (semiclassical) energy. 
Our approximation scheme thus reproduces the correct value of
$\xi$ as well as the corresponding value of the ground-state energy.
\\
The ground-state energy eigenvalue can also be obtained
once the Hamiltonian operator (\ref{BHM}) is linearized via the 
standard procedure $n_j^2 \approx 2\nu n_j -\nu^2$, which is 
realiable for $t \gg U$. This yields the diagonalized Hamiltonian
\be
H_{_{BH}} \simeq -U N_s \nu^2
+\! \sum_k [2U \nu- \mu - t\, g(k) ]\, b^+_k b_k \, ,
\label{BOGOL}
\ee
when the operators
$b_k = {N_s}^{-1/2} \Sigma_j a_j \exp [i {\tilde k} j]$
of the $k$ modes are used.
We have introduced $g(k)= \Sigma_{r} cos(k_r)$ 
($r \in [1,D] $ on a $D$--square lattice) where $k_r$
is the $r$-th component of $\vec k$. Hamiltonian (\ref{BOGOL}) clearly
shows that its lowest eigenvalue is obtained through the depletion
of any mode $k \ne 0$ (boson condensation in the state with $k=0$).
As a consequence of the consistency condition
$\langle n \rangle \equiv \nu$ ($\Rightarrow N =N_s \nu $), once more
the energy is minimized by $\nu \equiv (\mu + qt/2)/(2U)$.  

A comparison with the quantum ground-state energy --known exactly
in the case $t/U =0$-- is important as well. For $\mu/U \in I(n)$ 
the eigenvalues of $H_{_{BH}}$ with integer filling $n= N/N_s$,
$E(\{n_j\})= \Sigma_j [U n_j^2-(\mu +U) n_j]$ reach their minimun
value $E_* =-U N_s\, n^2$ for $n_j =n$ and $\mu/U =2n-1$.
The on-site energy $\epsilon_* =-U|z_*|^4$ (see (\ref{ZSC}))
is found to attain exactly its quantum counterpart
$E_* /N_s =-U n^2$ in the limit $t/U \to 0$, $\mu /U \to 2n$,
namely at the point of $I(n)$ representing its top.

The second observation pertains the action $\cal S$ that represents
the phase of the macroscopic state $| \Phi \rangle$.
It raises a special interest since it is itself a macroscopic
quantity and thus is viable to experimental observations. In the
following we compare $\cal S$ with ${\cal S}_{mf}$ ({\it i.e.}
$\cal S$ in our TDMFT) as well as ${\cal S}_{mf}$ in the MI with
${\cal S}_{mf}$ in the SF.
\\
When Eqs. (\ref{SEE}) are inserted in (\ref{action}) then
$\dot {\cal S}$ reduces to
\be
{\dot {\cal S}} = U \Sigma_j |z_j|^4 \;,
\label{ANGSF}
\ee
where the explicit form of $z_j(\tau)$
is known only once Eqs.~(\ref{SEE}) have been really solved.  
\\
Inserting Eqs.~(\ref{MFE}) instead of ~(\ref{SEE}) in (\ref{action}),
and replacing $\cal H$ with its mean field version ${\cal H}_{mf}$,
involves
\be
{\dot {\cal S}_{mf}} = \Sigma_j \left [ U |z_j|^4 +
t( {\bar{z}}_j \psi_j +{\bar {\psi}}_j z_j) - 2t|\psi_j|^2
\right ] \, .
\label{ANG}
\ee
If $z_j \approx \psi_j$, then  ${\dot {\cal S}}_{mf}$ and
$\dot {\cal S}$ have essentially the same form. Such macroscopic
quantities may actually coincide at the low temperature regime if
the dynamics of both ${\cal H}_{mf}$ and ${\cal H}$ have solutions
characterized by $|z_j(\tau)|^2 \approx const$.
Within the present TDMFT ${\dot {\cal S}}_{mf}\equiv \dot {\cal S}$
since the condition  $\eta_i {\bar \eta}_j \approx 0$ implies
$z_j \approx \psi_j$; furthermore, $|z_j(\tau)|$ is strictly
time--independent.

In the MI the requantization rule~(\ref{REQUANT}) must be used.
The frequency ${\dot {\cal S}_{mf} }$ is obtained from (\ref{ANG}) by
inserting~(\ref{REQUANT}). ${\cal S}_{mf}$ reads as
\be
{\dot {\cal S}_{mf}} =  U N_s n^2 \,\,\, .
\label{SMI}
\ee
A transition that changes the filling from $n$ to $(n+1)$ involves a
change of the phase frequency amounting to $U N_s (2n+1)$.
The action density in the superfluid phase reads:
${\dot {\cal S}_{mf}} =  U N_s |z|^4 $, which
compared with the corresponding formula~(\ref{SMI}) shows that
the frequency is not quantized. In the SF phase transitions between
different configurations (different values of the filling)
occur continously.
%
%
\section{spin coherent state picture of XXZ model}
\noindent
Spin coherent states (SPS) $|\xi \rangle$ are defined as
\be
|\xi \rangle := D_S(\xi) e^{\xi S^+}{} |S, -S\rangle \, ,
\label{SCS}
\ee
where $S^+$ is the raising operator of the angular momentum
algebra $su(2)$ generated by $S_z$, $S_x = (S^+ +S^-)/2$, and
$S_y =(S^+ - S^-)/2i$ which fulfil the standard commutators
\be
[S^z, S^{\pm}] = \pm S^{\pm}2 \quad, \quad [S^+, S^-] = 2S_z \;.
\label{SBR}
\ee
$D_S(\xi) \doteq  1/(1+|\xi|^2)^S$ represents a normalizing factor,
whereas $|S,-S \rangle$, the so-called maximum weight vector,
satisfies the equation $S^-|S, -S\rangle =0$. The action of $S^{\pm}$
$$
S^{\pm} |S,m\rangle = \sqrt{(S \mp m) (S \pm m+1)} |S,m \pm 1\rangle
$$
is represented on the standard basis $\{ |S, m \rangle; |m|<S \}$ the
vectors of which obey the secular equation $S^z |S,m\rangle = m |S,m
\rangle$.
Making explicit the action of $S^+$ in (\ref{SCS}) supplies 
the spanned form
$$
| \xi \rangle =  D_S(\xi)
\sum^S_{m=-S} C_{m}(S) \xi^{m-S}|S, m \rangle
$$
of $|\xi \rangle$, with $C_{m}(S) \doteq \sqrt{(2S)!/(S-m)!(S+ m)!}$.
Based on such formula one is able to calculate the expectation values
\be
L_z = \langle S^z \rangle = S \frac{|\xi|^2 -1}{ |\xi|^2 +1 } \; ,
\label{LZ}
\ee
\be
L^* = \langle S^+ \rangle = S \frac{2 \xi}{|\xi|^2 +1} \; ,
\label{LR}
\ee
where
$ \langle \bullet \rangle \doteq \langle \xi| \bullet |\xi \rangle$,
and therefore to reconstruct the sphere equation
$(L^z)^2 +(L^x)^2 +(L^y)^2 = S^2$, where $L = L^x -iL^y= (L^*)^*$
for the classic spin $(L^x ,L^y ,L^z)$. Such an equation, in turn,
can be viewed as the classic counterpart
of the (quantum) Casimir equation $(S^z)^2 +(S^x)^2+(S^y)^2 =S(S+1)$
for $(S^x,S^y,S^z)$ in terms of SCS picture and illustrate the
semiclassical content thereof. Upon introducing the macroscopic
wavefunction
\be
| \Phi \rangle \doteq  e^{i{\cal S}/ \hbar}| \xi \rangle
\label{STS}
\ee
where the trial state $|\xi\rangle {\cdot =}\otimes_i |\xi_i \rangle$
and $| \xi_i \rangle$ is the SCS for the i-th spin
$( S_i^x ,S_i^y ,S_i^z )$, then one easily constructs
the TDVP semiclassical dynamics relative to
spin Hamiltonian (\ref{SHAM1})
\begin{eqnarray}
H_{{_S}} = - h\sum_{i} \,(S^{z}_{i} +S ) &+& U\sum_{j}
 \, (S^{z}_{j} +S )( S^{z}_{j}+S-1 ) \nonumber \\
&-& \frac{E_{_S} }{2} \sum_{\langle {i} , {j} \rangle}
\left (  S^{+}_{i} \; S^{-}_{j} +
 S^{+}_{j} \; S^{-}_{i} \right ) \; ,
\nonumber
\end{eqnarray}
(now assumed to be constituted by spins with $S > 1/2$),
by proceeding along the same lines of appendix A. The resulting
Hamiltonian ${\cal H}_{_S} := \langle \xi | H_{_S} |\xi \rangle$
reads
\begin{eqnarray}
{\cal H}_{S} &=& -h_* \sum_{i} \, L^{z}_{i} 
+U \sum_i \,[(1- 1/2S)(L^{z}_{i} )^2 + S/2] \nonumber \\
&-& \sum_i S[US +h_*] 
- \frac{E_{_S}}{2} \sum_{\langle {i}, {j} \rangle}
\left (L^{*}_{i} \; L_{j} + L^{*}_{j} \; L_{i} \right ) \; ,
\label{SSH}
\end{eqnarray}
where $h_* =h-U(2S-1)$, and we have used
the fact that, via a nontrivial calculation, one finds
$\langle (S^z)^2 \rangle =(1- 1/2S)(L^{z}_{i})^2 +S/2$.
Furthermore, stationarizing the action
${\cal S} =
\int dt \langle \xi |(i\hbar \partial_t -H_{_S} )|\xi \rangle$
provides the equations of motion for the variables
$L^*_i$, $L_i$ (see Eqs.~(\ref{SEESPIN})),
where $L^z_i$ is depending on $L_i$, $L^*_i$ via the constraint
introduced above for the spin expectation value components
$(L^z_{i})^2 + |L_i|^2 \equiv S^2$. Once the brackets
\be
\{ A , B \}
= \sum_j {\frac{(1+|\xi_j|^2 )^2}{2Si \hbar}}  \left[
{\frac {\partial A }{\partial \xi_j }}
{\frac {\partial B }{\partial \xi^*_j  }}
- {\frac {\partial B }{\partial \xi_j }}
{\frac {\partial A }{\partial \xi^*_j }} \right]
\label{ALG}
\ee
have been defined, one can easily check that
$\{ L^*_j , L_j\} = 2 L^z_j/ i\hbar$,
$\{ L^z_j , L_j^* \} = L_j^*/ i\hbar$,
and $\{ L^z_j , L_j \} = -L_j/ i\hbar$
consistently with (\ref{SBR}), while the dynamical equations
issued from the TDVP can be recovered as well from ${\cal H}_{_S}$.
Equivalently, the alternative form of the above Lie-Poisson
brackets
\be
\{ A , B \}
=  {\frac{1}{\hbar}} \sum_j  \left[
{\frac {\partial A }{\partial \phi_j }}
{\frac {\partial B }{\partial L^z_{j}  }}
- {\frac {\partial B }{\partial \phi_j }}
{\frac {\partial A }{\partial L^z_{j}  }} \right]
\label{AAB}
\ee
can be reconstructed from (\ref{ALG}) when expressing $L_i$'s as
\begin{equation}
L_i = \sqrt{S^2 - (L^z_i)^2} \, e^{i \phi_i}.
\label{AZ}
\end{equation}
through the action-angle variables $\phi_i$, $L^z_i $. This
fact states at the classical level the equivalence between
the HPR and the VR introduced in Sec. III. 
%
%
%
%
\section{Local phase dynamics of SHM Hamiltonian}
This appendix is devoted to calculate explicitly
the form assumed by the spin Hamiltonian in proximity of
the ground-state configuration in order to show how 
weakly excited states mimic the dynamics of
the (classic) phase model. Upon recalling that the
ground-state configuration is characterized by
$$
{\cal L}_0 \equiv h^*/(2U+qE_{_S}) \, , \, \phi_i =\phi_j \, ,
$$
at each site $j$, considering the approximation
$$
L_j L_i^* +L_i L_j^* \simeq 
$$
$$
\simeq
2g \,cos(\phi_i -\phi_j) \times
\left [1 - \frac{{\cal L}_0}{g}
(P_i +P_j) - f(P_i,P_j) \right ]\, ,
$$
where $P_j =L_j^z -{\cal L}_0$,
$g =S^2 -{{\cal L}_0}^2$, and
$$
f(P_i,P_j) := \left [{\frac{ {{\cal L}_0 }^2}{2g^2} }
(P_i - P_j)^2 + {\frac{P_i^2 +P_j^2}{2g}} \right ]\, ,
$$
leads to rewrite spin Hamiltonian (\ref{XXZM}) as
\begin{eqnarray}
&{\cal H}_{S}& \, \simeq C+ \sum_i (U\,P^2_{i} +
U {{\cal L}_0 }^2 - h_* {\cal L}_0 )
\nonumber \\ 
&-& g E_{_S} \sum_{\langle {i}, {j} \rangle} \,f(P_i,P_j) 
-g E_{_S} \! \sum_{\langle {i}, {j} \rangle} \!
cos(\phi_j -\phi_i) \, .
\nonumber \\ 
\end{eqnarray}
Decoupling $f(P_i, P_j)$ from the cosine term in the latter formula
relies on the fact that $(\phi_j -\phi_i)^2 \times f(P_i, P_j)$ is
fourth order.
The resulting model exhibits the QPM structure even if, within the
present approximation scheme, the condition $|P_j| \ll {\cal L}_0$
only concerns the spin dynamics. The geometry of the sphere (the
spin configuration space) is involved instead when one imposes
$\beta =|m|/S \ll 1$ ($m$ is the quantum number corresponding to
${\cal L }_0$) in order to make explicit the (local) cylinderlike
geometry characterizing the QPM in the equatorial region.
\vspace{-0.5cm}
\end{appendix}

\end{multicols}

\newpage

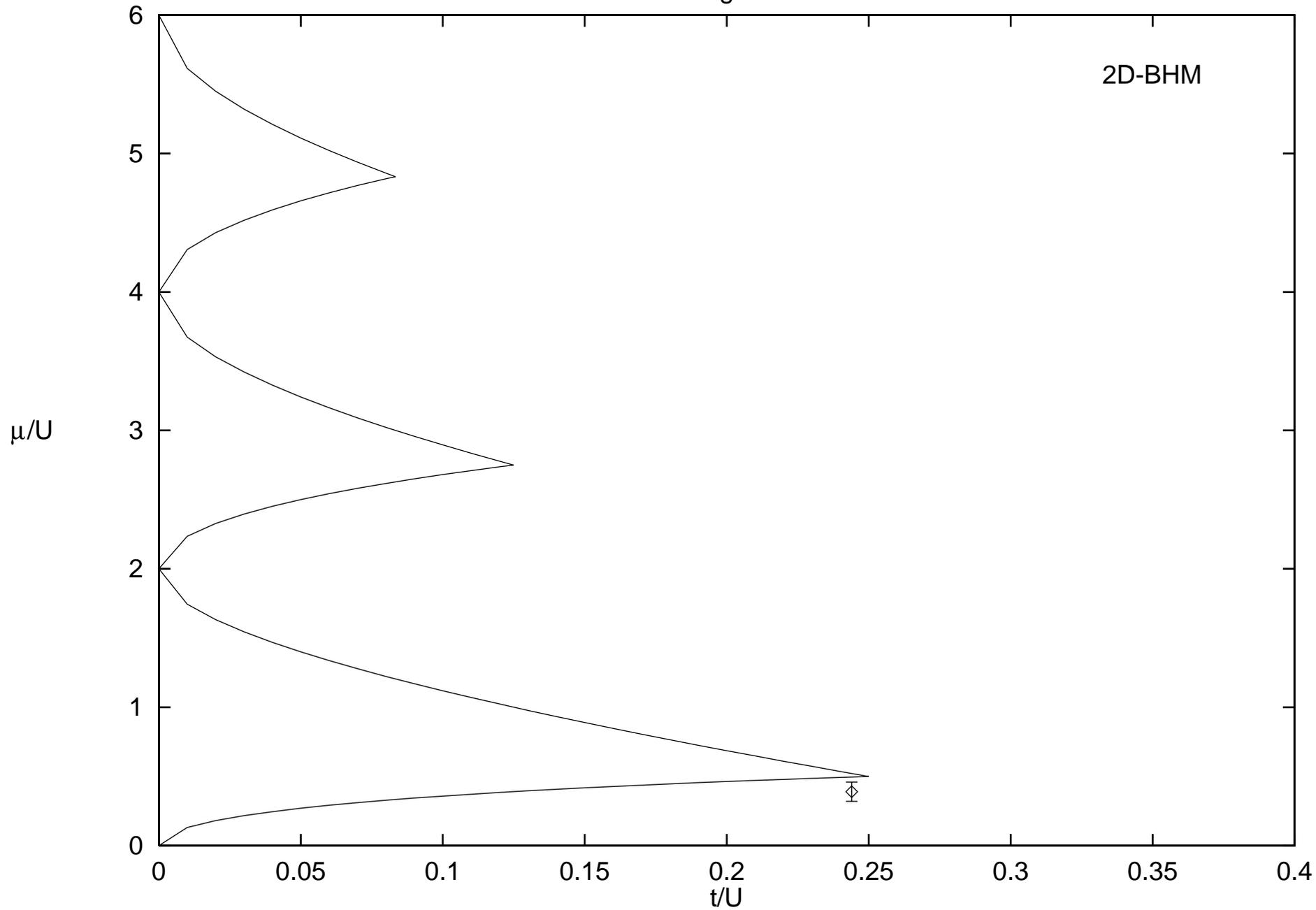
\begin{figure}
\caption{The Phase Diagram of the  $BHM$ for $D=2$ (continuous curve).
The error box indicates the QMC tricritical point
obtained by {\it Krauth and Trivedi} in Ref.[9].
For $n=1$, $(t_{c}/U)=0.25$ while
QMC gives $(t_{c}/U)=0.244 \pm 0.002$ and SCPE provides
$(t_{c}/U)=0.272$.}
\label{BHMD2}
\end{figure}
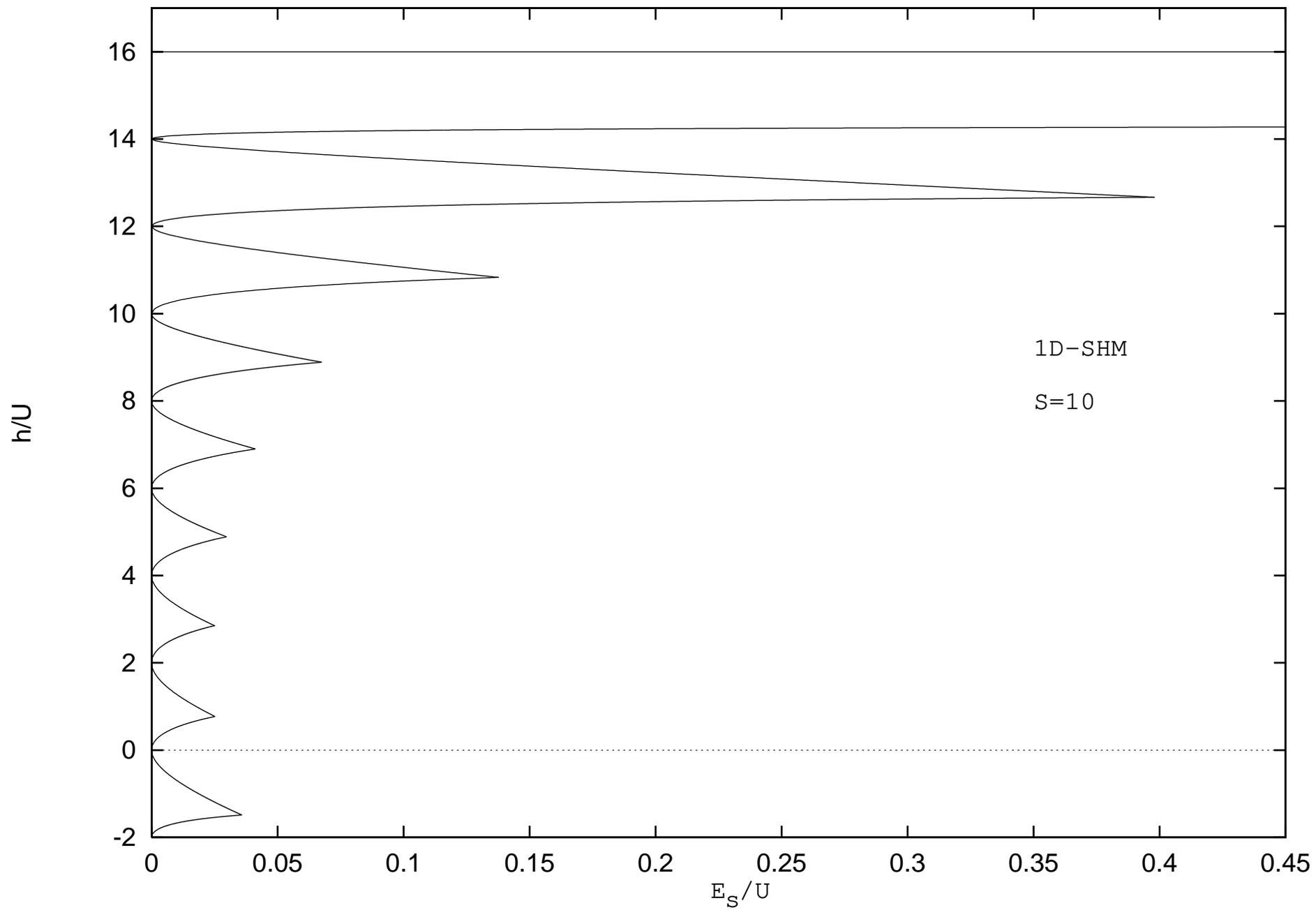
\begin{figure}
\caption{The phase diagram of the SHM for $D=1$ and $S=10$.
Inside the lobes the ground state is paramagnetic. Elsewhere the system 
is in a canted phase. 
The lobes are obtained for $m$ ranging from $m=-S$ (corresponding 
to $-2<h/U<0$ in the figure) to 
$m=1$ (the upst lobe). The coordinates $E_{_S}/U $
of the first two lobes' (corresponding to the $-2<h/U<2$ lobes) 
tricritical points decrease; they are 
``almost'' constant for $0<h/U<4$ (named ``inversion region'' in the text).
The SF phase is progressively reduced for increasing $m$ ({\it id. e.} 
such that $4<h/U<16$).}
\label{S-XXZ}
\end{figure}
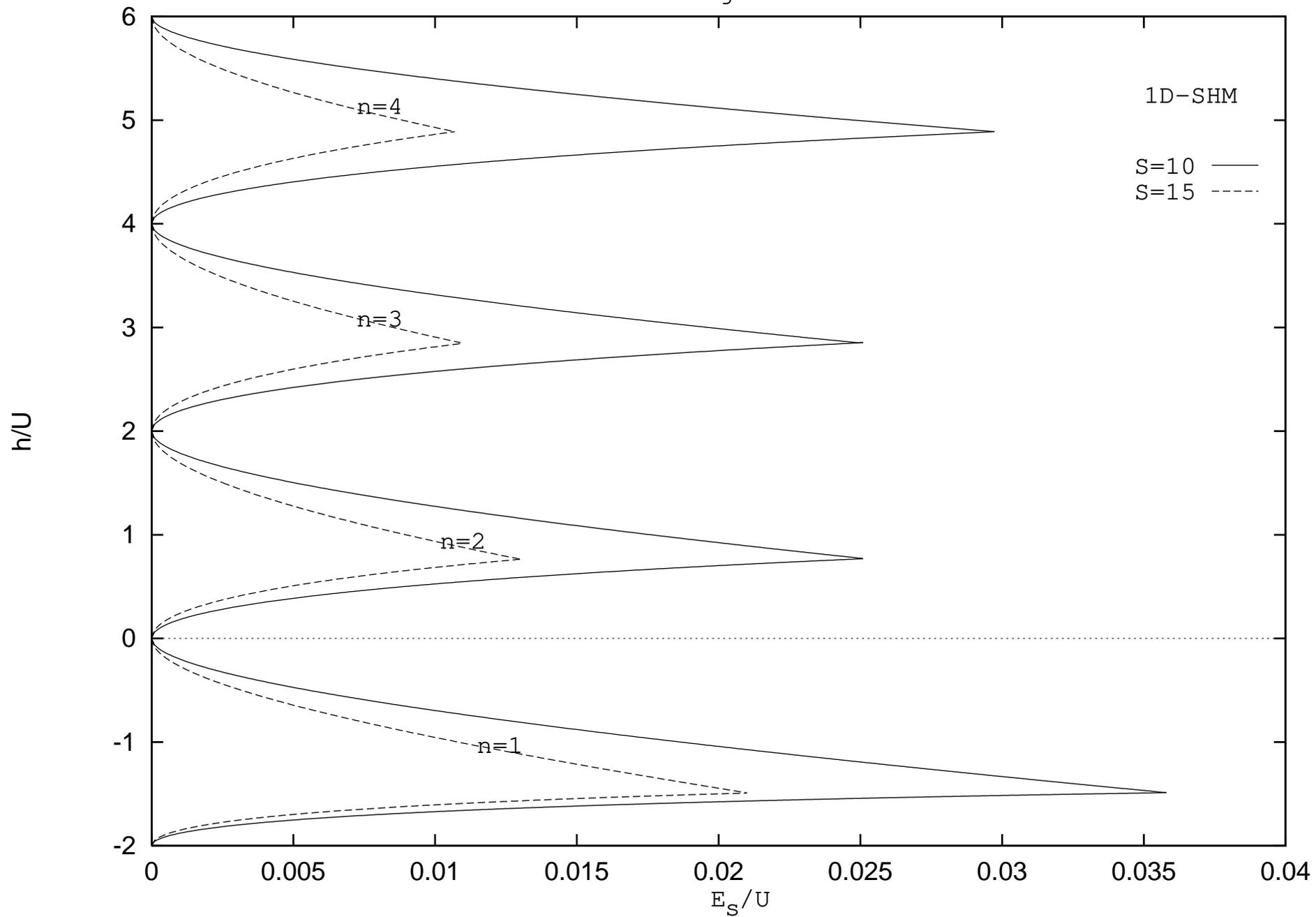
\begin{figure}
\caption{The phase diagram of the one dimensional $S$--XXZ model for $S=10$
and $S=15$. The lobes are obtained for $m$ ranging $(-S\dots 1)$; in 
the figure, $n=m+S$ ranges in $\{1\dots 4\}$.
We note that the SF phase is enlarged for increasing $S$.}
\label{S10vsS15}
\end{figure}
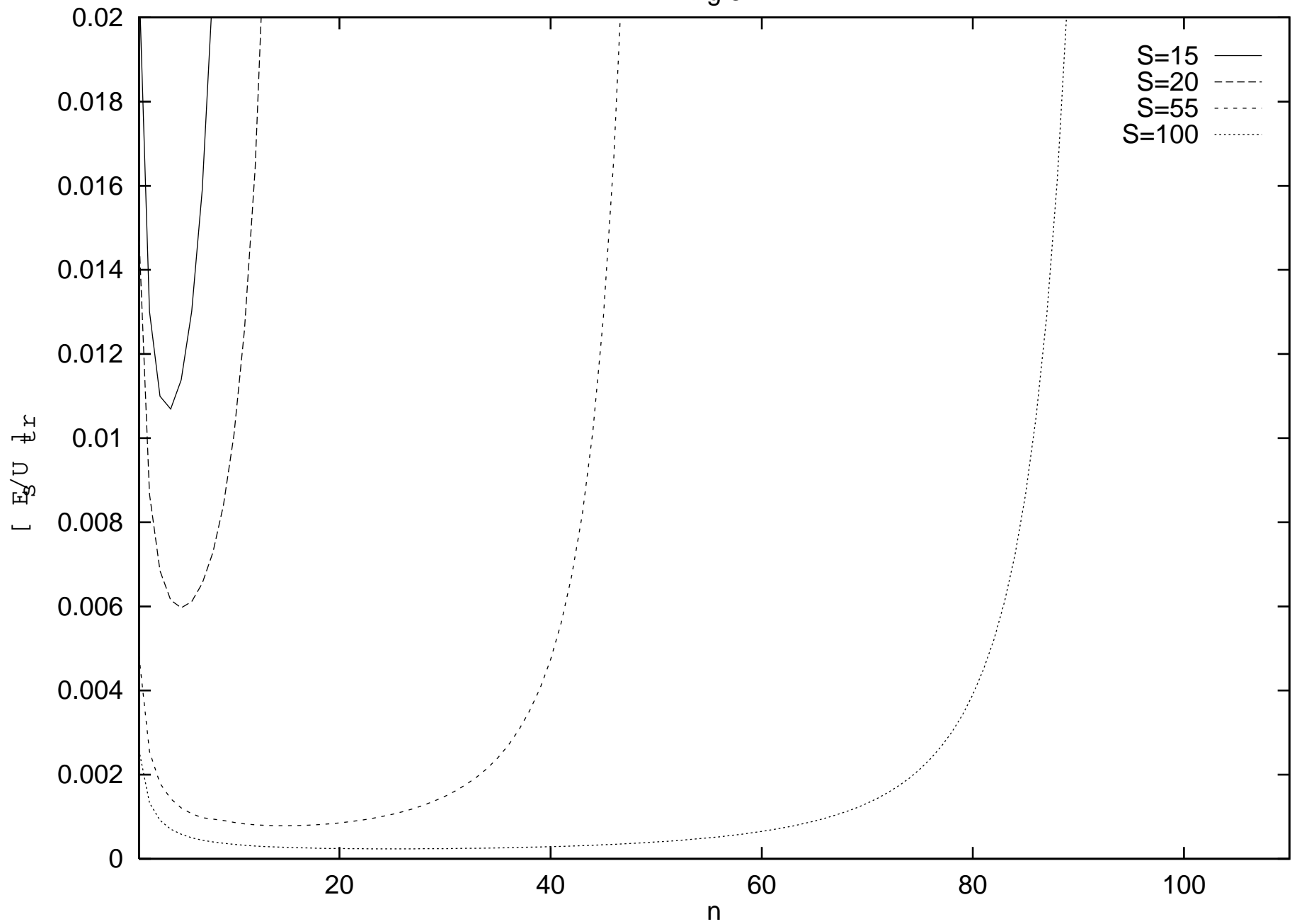
\begin{figure}
\caption{The behaviour of the tricritical points $(E_{_S} /U)_{tr}$ as 
function of $n$ for different values of $S$. 
We note that the region in which $(E_{_S} /U)_{tr}$ is 
almost constant (named inversion region in the text) becomes wider 
for increasing $S$. Whithin this region the SHM's dynamics can 
be considerd as pure phase dynamics. This behaviour is named ``QPM--like 
behaviour'' in the text.}
\label{TIPS}
\end{figure}
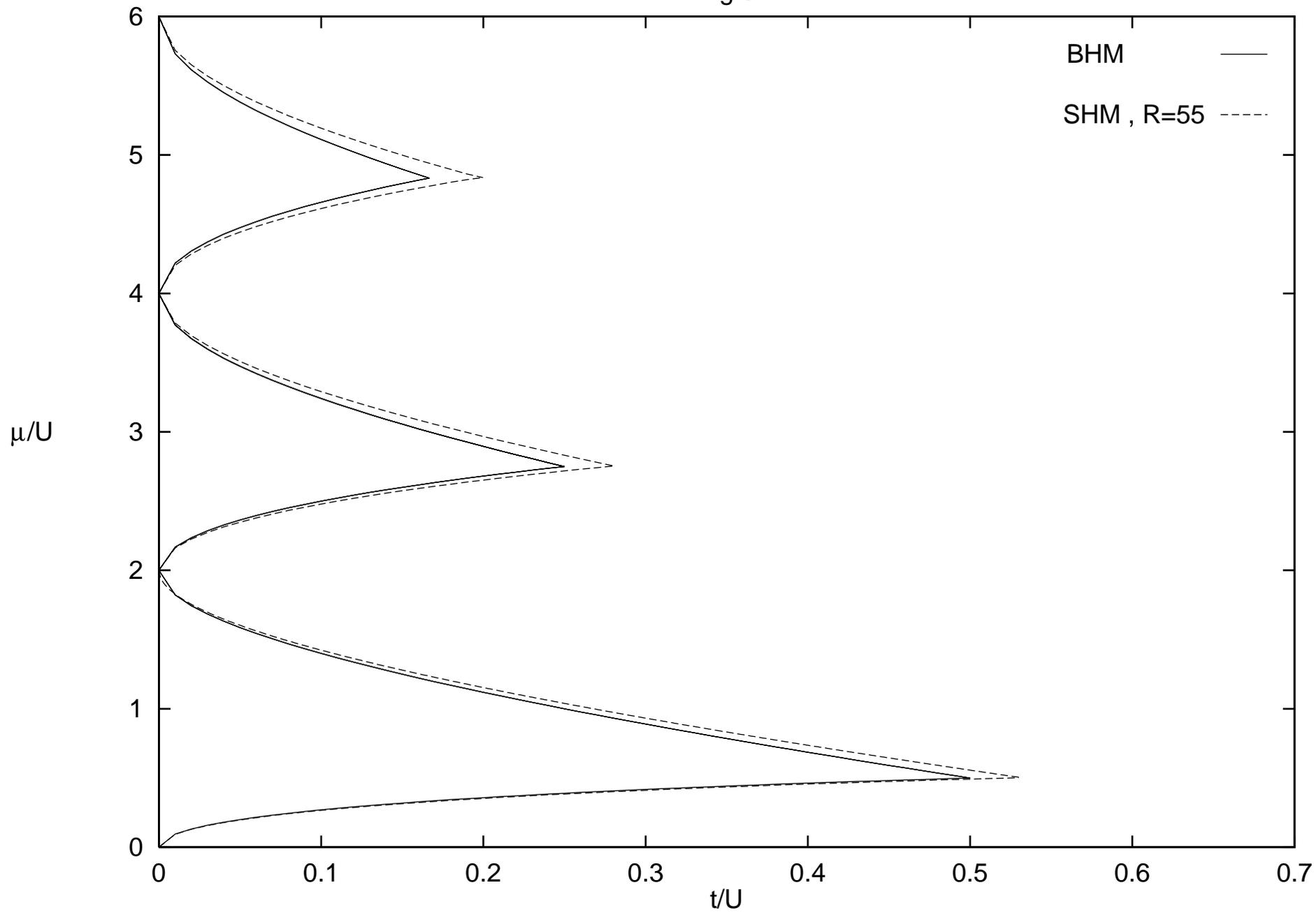
\begin{figure}
\caption{The phase diagram of the BHM compared with the phase diagram of the
SHM  for $S=55$. The two phase diagrams coincide 
considering the zero-th order in  $\alpha$ in formula~(\ref{ENEREXPA}) with $t=  2S E_{_S}$ and $h=\mu$}.
\label{XXZvsBHM}
\end{figure}
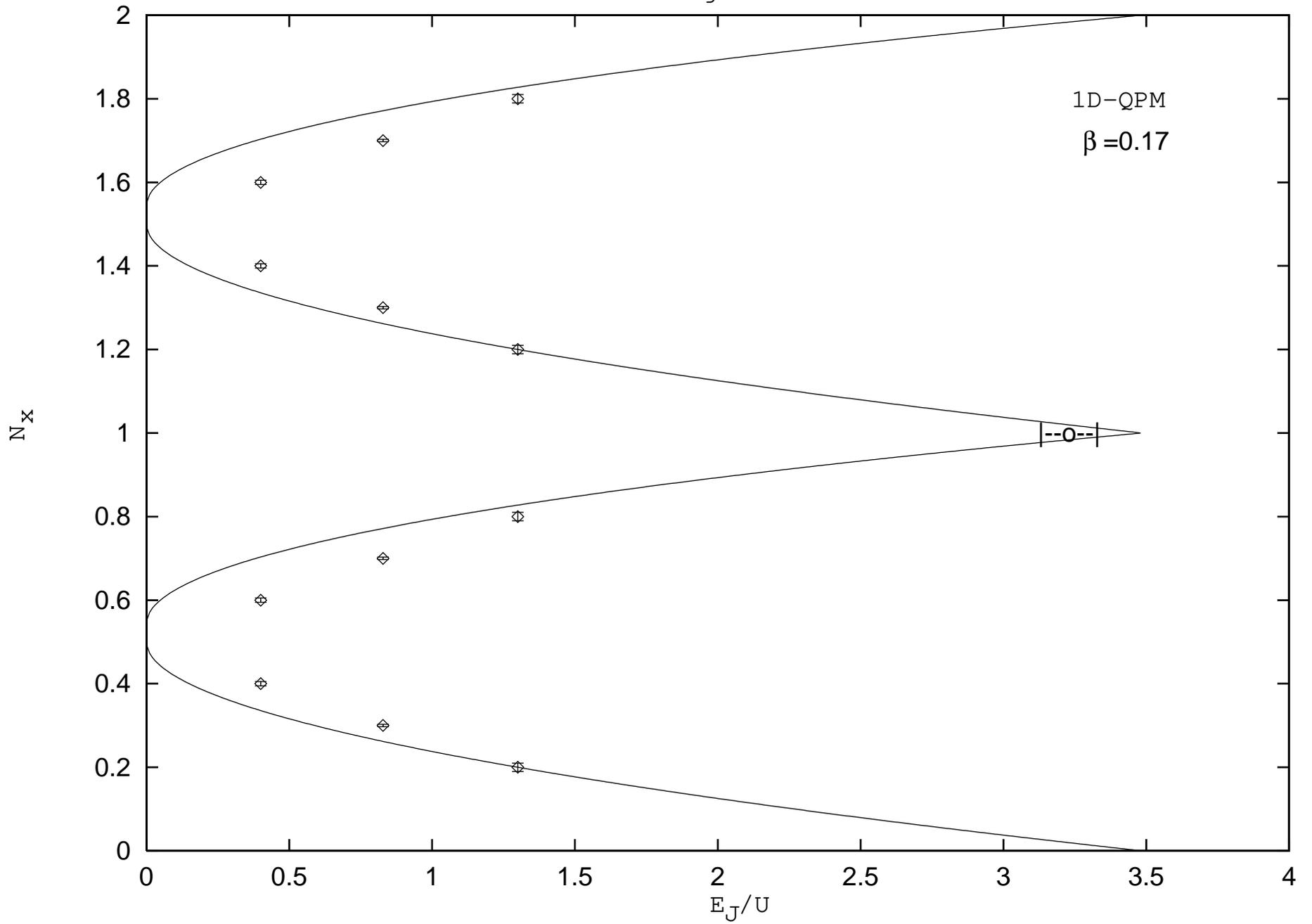
\begin{figure}
\caption{The phase diagrams of the SHM for
$S=55$ and $\beta=0.17$ in $D=1$. In Fig.6-a, the errorbars are 
the result of the QMC simulations of the QPM of Ref.[36].}
\end{figure}
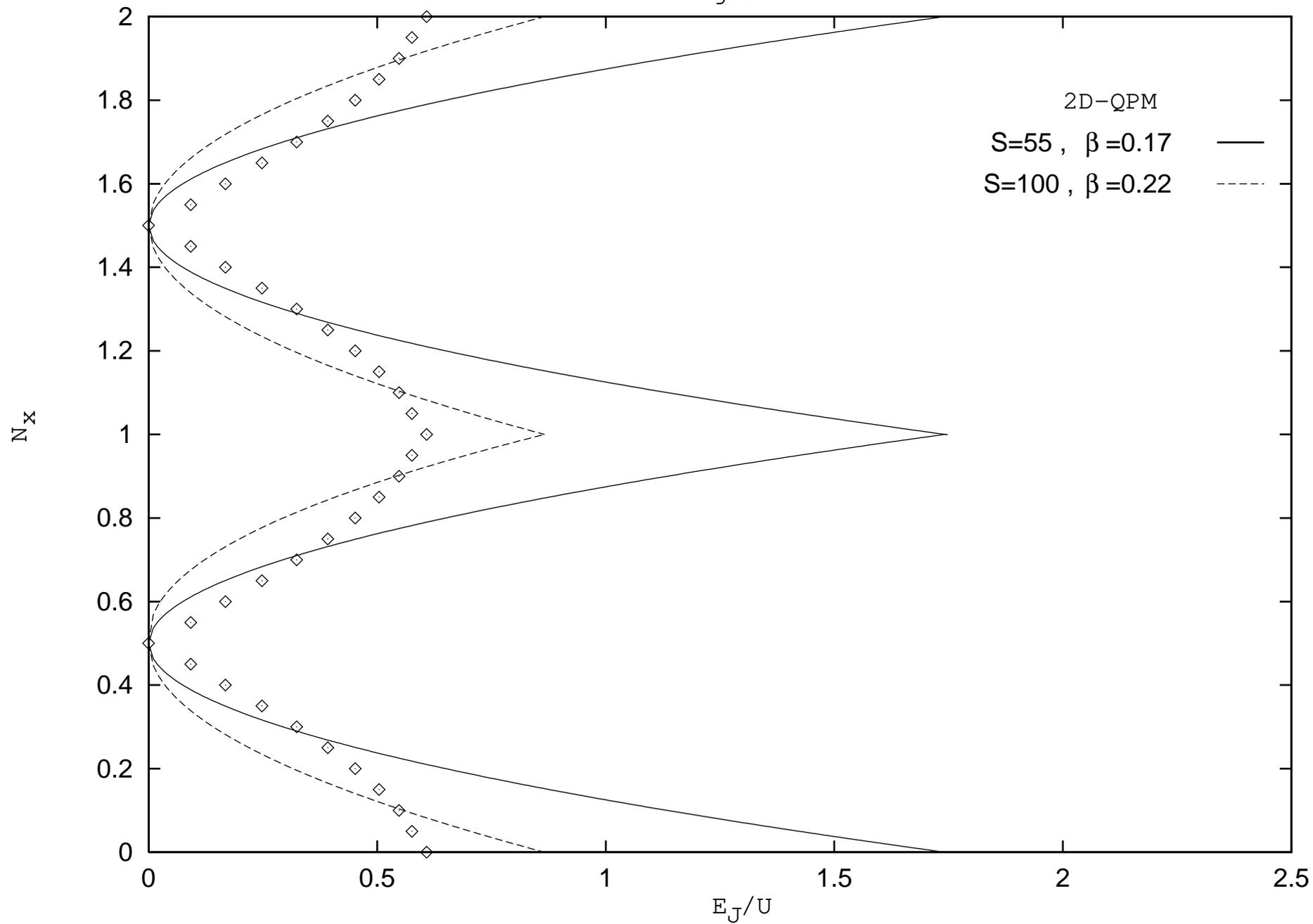
\begin{figure}
\caption{The phase diagrams of the SHM for
$S=55$ and $\beta=0.17$ and for $S=100$ and $\beta=0.22$ in $D=2$.
The boxes are result of QMC simulation of Ref.[37].}
\label{QPMPD}
\end{figure}


\begin{references}
\bibitem{NOTEFLUCT} The competition between the quantum fluctuation 
of $n_i$ and  $\Phi_i$ has a link with the energetic competition 
between the Coulomb and kinetic energies: the hopping processes,
which tend to delocalize the particles, induce fluctuations of $n_j$
thus reducing phase's fluctuations, and favouring the macroscopic
phase coherence. On the contrary, a finite electrostatic energy
tends to localize bosons, reduces the bosons number fluctuations, and
consistently prompts strong fluctuations of $\Phi_j$'s at each site.
\bibitem{NOTEFT} 
We note how the set of $z_j$ corresponds to the lattice version of
the order parameter field used in the standard scalar field theories
for superfluidity. 
In this respect the slow part of $z_j$ indeed should represent
the order parameter field of a superfluid medium when the
contributions relative to small time scale are negligible~\cite{AHNS}.
In fact, the usual canonical conjugation between the phase
of the superconducting order parameter and the total particle number
$N$ (see f.i.~\cite{TITI}) is inherited naturally by the TDVP effective
theory of the BHM.
In this picture, in fact, the low temperature dynamics is dominated
by the $z_j$ phases which is the condition whereby showing that
the effective action (the phase of the medium wavefunctions) is
canonically conjugated to $N$.
\bibitem{NOTECAN} We could define a hermitian  phase operator
$\Phi$, but it would not canonically conjugated to $n$.
\bibitem{NOTELIM1}In the limit, the spin projection along $z$
follows its representation.
\bibitem{NOTELIM2} In the limit, the $z$--spin's
projection is kept finite. Only after the limit, $S^z_i$ range on a
infinite interval.
\end{references}
\end{document}